\documentclass[10pt,twocolumn,letterpaper]{article}

\usepackage{cvpr}              % To produce the CAMERA-READY version
\usepackage[dvipsnames]{xcolor}

\definecolor{cvprblue}{rgb}{0.21,0.49,0.74}
\usepackage[pagebackref,breaklinks,colorlinks,citecolor=cvprblue]{hyperref}
\usepackage{booktabs}       % professional-quality tables
\usepackage{multirow}

\def\paperID{3236} % *** Enter the Paper ID here
\def\confName{CVPR}
\def\confYear{2024}

\title{Boosting Neural Representations for Videos with a Conditional Decoder}

\author{Xinjie Zhang\textsuperscript{1,2}\thanks{Work was done when Xinjie Zhang interned at SenseTime Research} \quad \quad
Ren Yang\textsuperscript{2}\thanks{Corresponding author} \quad \quad
Dailan He\textsuperscript{3} \quad \quad
Xingtong Ge\textsuperscript{4} \\
Tongda Xu\textsuperscript{5} \quad \quad
Yan Wang\textsuperscript{5} \quad \quad
Hongwei Qin\textsuperscript{2} \quad \quad
Jun Zhang\textsuperscript{1}\textsuperscript{\textdagger}\\
$^{1}$The Hong Kong University of Science and Technology \quad $^{2}$SenseTime Research \\ $^{3}$The Chinese University of Hong Kong \quad $^{4}$Beijing Institute of Technology \\
$^{5}$Institute for AI Industry Research (AIR), Tsinghua University\\
\texttt{\footnotesize xzhangga@connect.ust.hk, r.yangchn@gmail.com, hedailan@link.cuhk.edu.hk, xingtong.ge@bit.edu.cn}\\
\texttt{\footnotesize x.tongda@nyu.edu, wangyan202199@163.com, qinhongwei@sensetime.com, eejzhang@ust.hk}
}

\begin{document}
\maketitle
\begin{abstract}
Implicit neural representations (INRs) have emerged as a promising approach for video storage and processing, showing remarkable versatility across various video tasks. However, existing methods often fail to fully leverage their representation capabilities, primarily due to inadequate alignment of intermediate features during target frame decoding. This paper introduces a universal boosting framework for current implicit video representation approaches. Specifically, we utilize a conditional decoder with a temporal-aware affine transform module, which uses the frame index as a prior condition to effectively align intermediate features with target frames. Besides, we introduce a sinusoidal NeRV-like block to generate diverse intermediate features and achieve a more balanced parameter distribution, thereby enhancing the model's capacity. With a high-frequency information-preserving reconstruction loss, our approach successfully boosts multiple baseline INRs in the reconstruction quality and convergence speed for video regression, and exhibits superior inpainting and interpolation results. Further, we integrate a consistent entropy minimization technique and develop video codecs based on these boosted INRs. Experiments on the UVG dataset confirm that our enhanced codecs significantly outperform baseline INRs and offer competitive rate-distortion performance compared to traditional and learning-based codecs. Code is available at \url{https://github.com/Xinjie-Q/Boosting-NeRV}.
\end{abstract}

\section{Introduction}
\label{sec:intro}

\begin{figure}[t]
 \centering
  \subfloat
  {\includegraphics[scale=0.3]{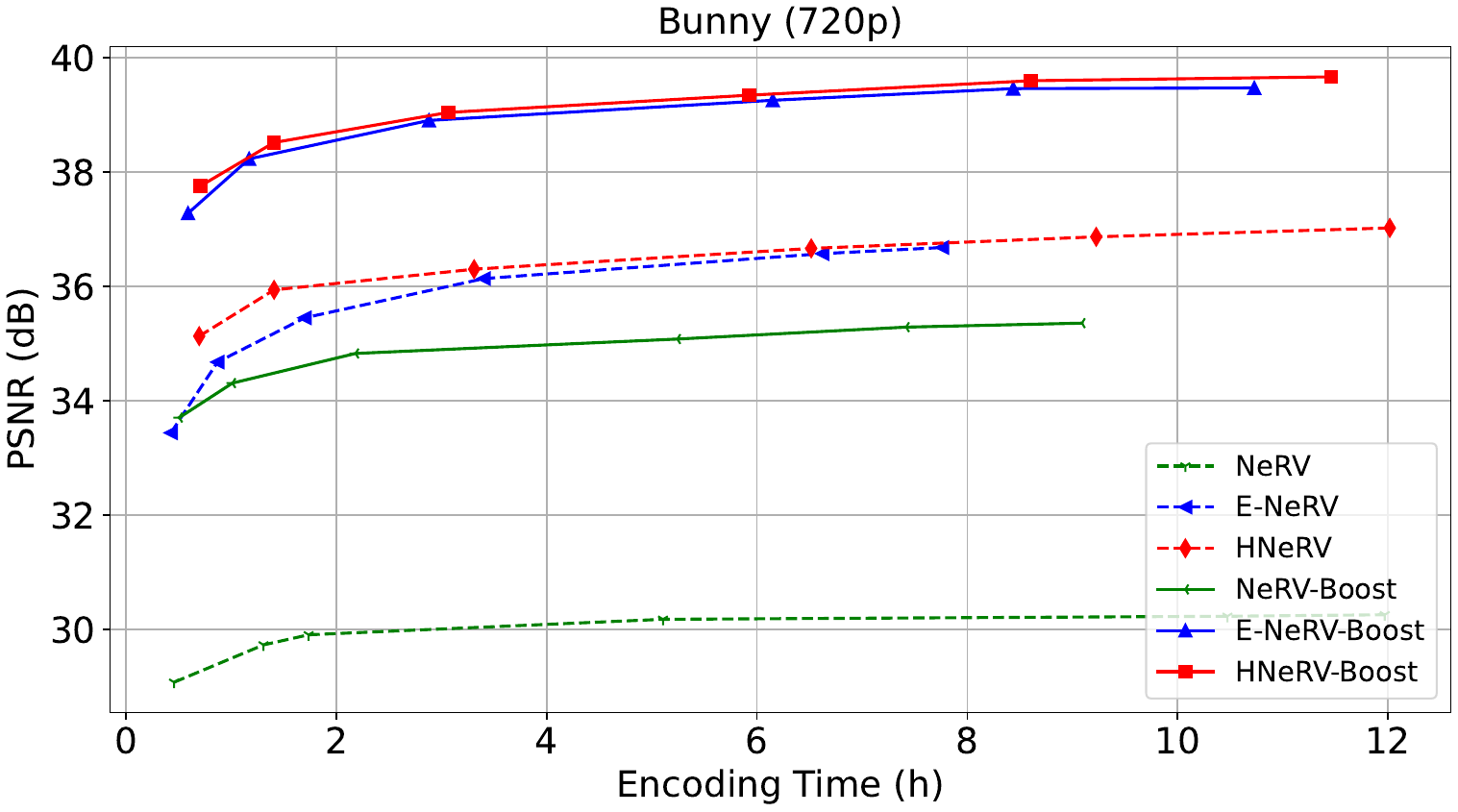}}\\
  \subfloat
  {\includegraphics[scale=0.3]{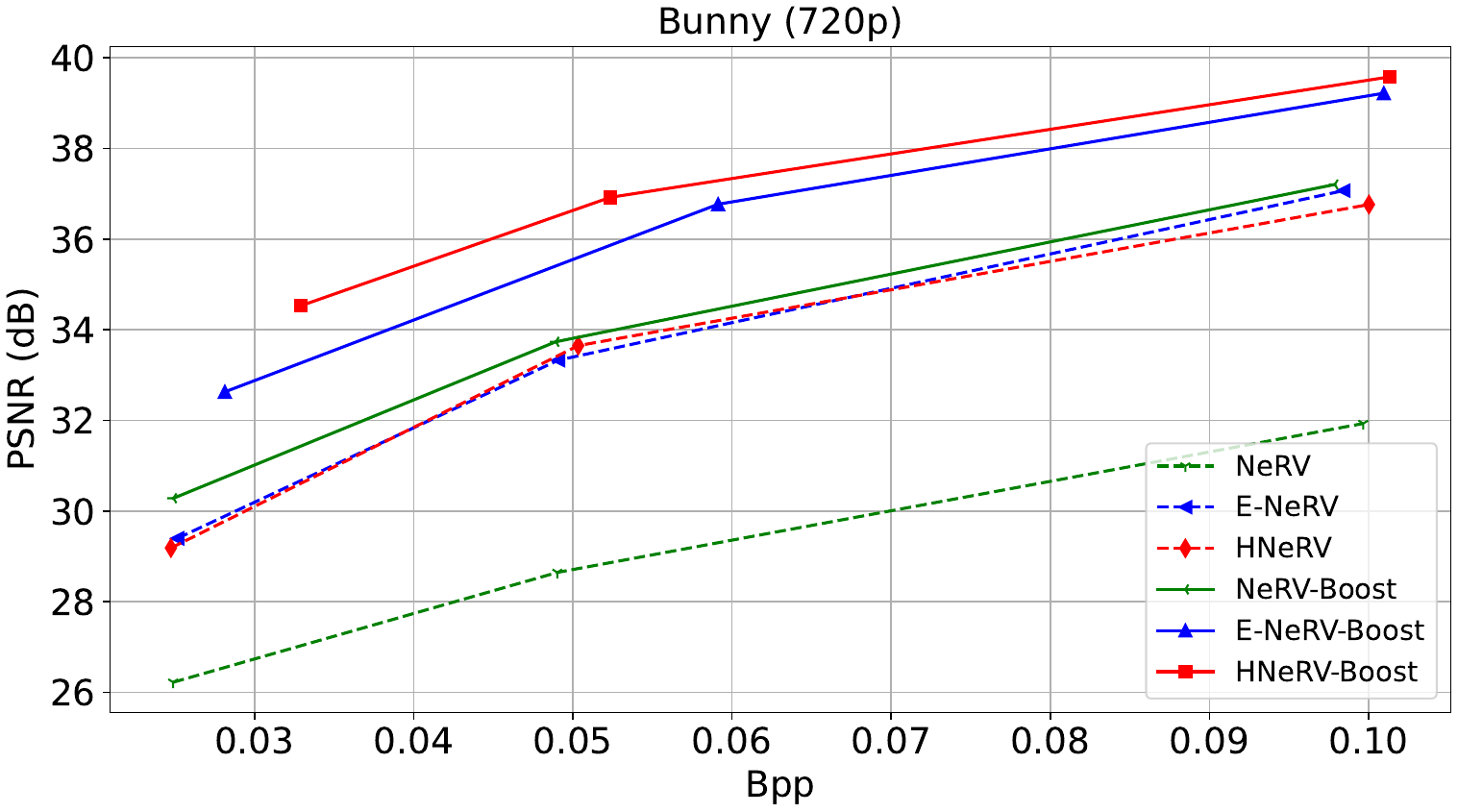}}
   \vspace{-0.3cm}
  \caption{Video regression with different encoding time under 1.5M model size (left) and video compression with different model sizes(right). Our boosted methods  achieve significantly better performance than the corresponding baselines.
  %Video regression with different model sizes under 300 epochs (top) and different epochs under 1.5M model size (bottom).
  }
  \label{fig:buuny_regression}
  \vspace{-0.6cm}
\end{figure}
Implicit neural representations (INRs) are gaining widespread interest for their remarkable capability in accurately representing diverse multimedia signals, including audios \cite{sitzmann2020implicit, szatkowski2022hypersound}, images \cite{muller2022instant, saragadam2023wire}, and 3D scenes \cite{mildenhall2021nerf, park2021nerfies, pumarola2021d}. They typically use a compactly parameterized neural network to learn an implicit continuous mapping that translates coordinates into target outputs (\eg, RGB values, density). This new-fashioned neural representation has opened up a plethora of potential applications, ranging from data inpainting \cite{xu2023revisiting, saragadam2023wire, li2023regularize} and signal compression \cite{strumpler2022implicit, zhang2022implicit, yang2023sci} to advanced generative models \cite{skorokhodov2021adversarial,skorokhodov2022stylegan}.

% Implicit neural representations (INRs) have recently attracted increasing attention due to their strong ability in fitting various multimedia signals, such as audios \cite{sitzmann2020implicit, szatkowski2022hypersound}, images \cite{muller2022instant, saragadam2023wire} and 3D scenes \cite{mildenhall2021nerf, park2021nerfies, pumarola2021d}. They typically use a compactly parameterized neural network to learn an implicit continuous mapping that maps coordinates to desired quantity of interest (\eg, RGB values, density). This new-fashioned neural representation has exhibited numerous potential applications including data inpainting \cite{xu2023revisiting, saragadam2023wire, li2023regularize}, signal compression \cite{strumpler2022implicit, zhang2022implicit, yang2023sci}, and generative models \cite{skorokhodov2021adversarial,skorokhodov2022stylegan}. 

Given their simplicity, compactness and efficiency, several studies have suggested applying INRs to video compression. Unlike traditional \cite{wiegand2003overview, sullivan2012overview} and recent neural \cite{lu2019dvc, lin2020m, li2021deep, sheng2022temporal} video codecs that rely on a complex predictive coding paradigm with separate encoder and decoder components, NeRV \cite{chen2021nerv} pioneers in representing videos as a function of the frame index $t$ and formulating video compression as model-based overfitting and compression. This innovative method significantly simplifies the encoding and decoding processes. Built on this paradigm, a series of subsequent works \cite{li2022nerv, chen2022cnerv, chen2023hnerv, zhao2023dnerv, lee2023ffnerv} have devoted to designing more meaningful embeddings to improve the quality of video reconstruction.

However, there are several vital limitations hindering the potential of existing implicit video representations. \textbf{Firstly}, when decoding the $t$-th frame, the identity information in most works only relies on the $t$-th temporal embedding. This approach often struggles to align the intermediate features with the target frame. Although a few works \cite{li2022nerv, bai2023ps, gomes2023video} introduce the AdaIN \cite{huang2017arbitrary} module to modulate intermediate features, this couples normalization and conditional affine transform. Its normalization operation might reduce the over-fitting capability of the neural network, resulting in limited performance gains (See \tableautorefname~\ref{table:ablation}). \textbf{Secondly}, while there are several studies \cite{li2022nerv, lee2023ffnerv, chen2023hnerv} in refining NeRV's upsampling block for a more streamlined convolutional framework, the impact of activation layers on the model's representational ability remains under-explored. 
\textbf{Moreover}, most previous methods rely on the L2 loss \cite{chen2023hnerv, zhao2023dnerv} or a combination of L1 and SSIM losses \cite{chen2021nerv, li2022nerv, chen2022cnerv, lee2023ffnerv} to overfit videos, but they often fail to preserve high-frequency information (\eg, edges and fine details within each frame), thereby degrading the reconstruction quality (See \tableautorefname~\ref{table:ablation}). \textbf{Finally}, most video INRs follow NeRV to employ a three-step model compression pipeline (\ie, pruning, quantization, and entropy coding) in the video compression task. Nevertheless, these components are optimized separately, which prevents INRs from achieving optimal coding efficiency. Although a few works \cite{gomes2023video, maiya2023nirvana} have explored the joint optimization of quantization and entropy coding, they face critical challenges arising from inconsistencies in the entropy models employed during both training and inference stages, leading to sub-optimal rate-distortion (RD) performance (See \tableautorefname~\ref{table:compression_comparision} and Fig.~\ref{fig:em}). Hence, we argue that there is great potential to boost the performance by overcoming the challenges we mentioned above.

To this end, we propose a universal boosting framework based on a conditional decoder to deeply explore the representation performances of existing video INRs. 
\textbf{Firstly}, we introduce a temporal-aware affine transform (TAT) module that discards normalization to better align intermediate features with the target frame. It is achieved by using a pair of affine parameters $(\boldsymbol{\gamma}, \boldsymbol{\beta})$ derived from temporal embeddings. We further incorporate a residual block with two TAT layers to facilitate both feature alignment and information retention. By strategically alternating upsampling blocks and TAT residual blocks, our conditional decoder significantly boosts the model's representation capabilities. 
\textbf{Secondly}, as shown in Fig.~\ref{fig:feature_maps}, we find that the GELU layer in the NeRV-like block activates only a limited number of feature maps. To address this issue, we introduce a sinusoidal NeRV-like (SNeRV) block to replace the GELU layer with a SINE layer to generate more diverse features. By using a small kernel size in the SNeRV block and placing more SNeRV blocks in later upsampling stages, we achieve a more balanced parameter distribution across the network, which helps improve the model's capacity.
\textbf{Moreover}, we integrate L1, MS-SSIM, and frequency domain losses as our optimization objectives during overfitting a video. This trio of loss functions promises to preserve intricate details in the reconstructed videos.
\textbf{Finally}, we advocate a consistent entropy minimization (CEM) technique based on a network-free Gaussian entropy model with tiny metadata transmission overhead, which not only ensures the consistency of training and inference, but also captures the interrelationships between elements in each weight or embedding to accurately estimate the probability distribution. 

In summary, our main contributions are three-fold: 
(1) We develop a universal boosting framework based on a novel temporal-aware conditional decoder to effectively improve the representation capabilities and accelerate the convergence speed compared to existing video INRs, shown in Fig.~\ref{fig:buuny_regression}.
% We develop a novel temporal-aware conditional decoder that incorporates the TAT residual block and SNeRV block. With a sophisticated loss function combining L1, MS-SSIM, and frequency domain losses, the proposed framework effectively boosts the representation capabilities of existing implicit video representations. 
(2) We further design a consistent entropy minimization scheme based on a parameter-efficient Gaussian entropy model to eliminate the discrepancy between training and inference.
(3) Extensive experimental results demonstrate that our boosted video INRs achieves a remarkable performance improvement against various baselines on multiple tasks, including video regression, compression, inpainting and interpolation. Comprehensive ablations and analyses demonstrate the effectiveness of each proposed component.

%unify the weight and embedding compression. Consequently, our scheme is exceptionally suited for any video INR compression, marking a significant stride in this domain. 

% In summary, our main contributions are three-fold: (1) We introduce a conditional decoder, consisting of our proposed SNeRV block and TAT residual block, to improve the representation ability of existing video INRs. A novel loss function that combines L1, MS-SSIM and frequency domain loss is introduced to capture high-frequency details. (2) For video compression, we further propose a parameter-efficient Gaussian entropy model to ensure the consistency between training and inference during entropy coding, meanwhile providing a more accurate estimation of the probability distribution for quantized weights and embeddings. (3) Our proposed method is suitable for different video INR models. Compared with original video INR models, our improved versions achieve competitive performances on multiple datasets and tasks (\ie, video regression, compression, inpainting and interpolation). Comprehensive ablations and analyses demonstrate the effectiveness of each proposed component.

\section{Related Work}
\label{sec:related_work}

\begin{figure*}[t]
    \centering
    \includegraphics[scale=0.8]{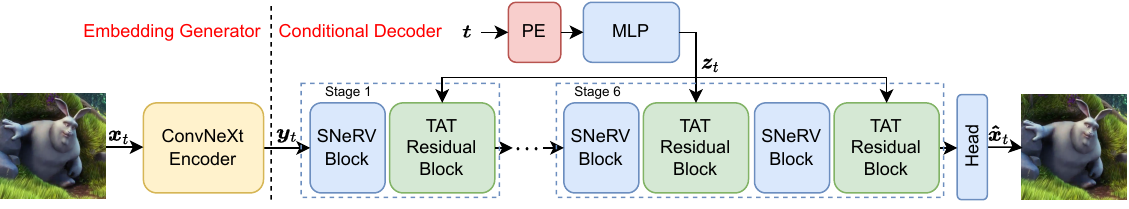}
    \caption{Our proposed HNeRV-Boost framework with the conditional decoder. The content-relevant embedding $\boldsymbol{y}_t$ expands its channel dimensions in stage 1 and upsamples in stages 2 to 6. The final three stages stack two SNeRV blocks with small kernel sizes to get fewer parameters, where the former upsamples features and the latter refines the upsampled features.}
    \label{fig:arch}
    \vspace{-0.5cm}
\end{figure*}

\noindent \textbf{Implicit Video Representation}. Recently, implicit video representations have engaged increasing interest due to their wide-ranging potential applications, such as video compression, inpainting and interpolation. Roughly, existing implicit video representations can be classified into two categories: (i) \textit{Index-based} methods \cite{chen2021nerv, li2022nerv, bai2023ps, maiya2023nirvana, gomes2023video} take content-independent time vectors and/or spatial coordinates as inputs. They only rely on the neural network to store all video information. (ii) \textit{Hybrid-based} methods \cite{chen2022cnerv, chen2023hnerv, zhao2023dnerv, lee2023ffnerv, tang2023scene} take content-relevant embeddings as inputs to provide a visual prior for the network, which reduces the learning difficulty of the model and thereby enhances its representation performance. These content-relevant embeddings can be derived from each frame using specific encoders \cite{chen2023hnerv, zhao2023dnerv, tang2023scene} or generated through random initialization and backpropagation \cite{chen2022cnerv, lee2023ffnerv}. 
In this paper, we focus on designing a universal conditional decoder framework to effectively boost the representation performance of existing video INRs, thus setting a new benchmark in the field.

\noindent \textbf{Network Conditioning.} Early methods in conditional feature modulation heavily rely on conditional normalization (CN). CN replaces the feature-wise transformation in normalization layers with affine parameters generated from external information, which has shown its effectiveness in applications like style transfer \cite{dumoulin2016learned, ghiasi2017exploring, huang2017arbitrary}, semantic image synthesis \cite{park2019semantic}, and denoising \cite{kim2020transfer}. A pivotal contribution by Perez \textit{et al.} \cite{perez2018film} demonstrates that it is possible to directly modulate intermediate features in a network without undergoing normalization. This technique has been widely used in super-resolution \cite{wang2018recovering, hu2019channel}, compression \cite{song2021variable}, and restoration \cite{wang2021towards}.
Instead of using CN to modify the distribution of intermediate features as in previous INR studies \cite{li2022nerv, bai2023ps, gomes2023video}, we introduce a conditional affine transformation without normalization to achieve a more precise alignment of intermediate features with the target frame, potentially improving the reconstruction quality of existing video INRs.

% \noindent \textbf{Network Conditioning.} Early conditional feature modulation methods focus on conditional normalization (CN) that replaces the feature-wise transformation in normalization layers with affine parameters dynamically generated from external information, which has proven effective in style transfer \cite{dumoulin2016learned, ghiasi2017exploring, huang2017arbitrary}, semantic image synthesis \cite{park2019semantic}, and denoising \cite{kim2020transfer}. Perez \textit{et al.} \cite{perez2018film} show that intermediate features in the network can be directly modulated without passing normalization. This operation has been widely used in super-resolution \cite{wang2018recovering, hu2019channel}, compression \cite{song2021variable}, and restoration \cite{wang2021towards}. In this paper, we aim to introduce conditional affine transformation without normalization to align intermediate features with the target frame instead of following previous studies \cite{li2022nerv, bai2023ps, gomes2023video} that use CN to alter the distribution of intermediate features. 

\noindent \textbf{INR for Video Compression.} As implicit neural networks generally fit videos using the model weights, it offers a novel perspective on video compression by translating it into model compression. Most video INRs \cite{li2022nerv, bai2023ps, lee2023ffnerv, kwan2023hinerv} follow the three-step compression pipeline of NeRV \cite{chen2021nerv}: (i) model pruning, such as global unstructured pruning, with fine-tuning to reduce the model size; (ii) post-training quantization or quantization-aware training to lower the precision of each weight; (iii) entropy coding to minimize the statistical correlation of coded symbols. Unfortunately, optimizing these components separately leads to sub-optimal coding efficiency. Thus, Gomes \textit{et al.} \cite{gomes2023video} and Maiya \textit{et al.} \cite{maiya2023nirvana} have applied entropy minimization \cite{balle2017end, balle2018variational, oktay2019scalable} to improve the video INR compression. During training, they estimate the bitrate of quantized weights using a small neural network to model each weight's distribution. But in the inference stage, this neural entropy model is replaced with either a context-adaptive binary arithmetic coder (CABAC) or an arithmetic coder using a fixed statistical frequency table. This switch in entropy models between training and inference leads to the sub-optimal RD performance. Moreover, these methods focus on compressing model weights, overlooking the significance of content-relevant embeddings in hybrid-based video INRs. To bridge these gaps, we propose a consistent entropy minimization technique to unify the weight and embedding compression. Consequently, our scheme is exceptionally suited for any video INR compression, marking a significant stride in this domain. 

\section{Method}
\label{sec:method}
As shown in Fig.~\ref{fig:arch}, our proposed boosted video representation architecture comprises two primary components: an embedding generator and a conditional decoder. The choice of an embedding generator depends on the specific video INR model. For clarity and ease of understanding, we take the hybrid-based representation model HNeRV \cite{chen2023hnerv} as an example. It is worth mentioning that our boosting framework can be easily generalized to other representation models (\eg, NeRV \cite{chen2021nerv}, E-NeRV \cite{li2022nerv}) by selecting appropriate embedding generators, with more details given in appendix. 

% In Section~\ref{subsec:overview}, we provide a comprehensive overview of our proposed boosting framework. The details of our conditional decoder are then elaborated in Section~\ref{subsec:cond_decoder}. Finally, in Section~\ref{subsec:compression}, we develop a novel INR compression method for video compression deployment.

% As shown in Fig.~\ref{fig:arch}, our proposed video representation architecture is composed of two parts, \ie, embedding generator and conditional decoder. The embedding generator depends on the specific video INR models. In this section, we select the embedding generator from the hybrid-based representation model HNeRV \cite{chen2023hnerv} for the sake of clarity and ease of understanding. It’s worth mentioning that our framework can also be generalized to other representation models (\eg, NeRV \cite{chen2021nerv}, E-NeRV \cite{li2022nerv}) when choosing different embedding generators. We first give an overview of the proposed framework in Section~\ref{subsec:overview}. Then we present the details of our proposed conditional decoder in Section~\ref{subsec:cond_decoder}. Finally, we develop a novel INR compression method in Section~\ref{subsec:compression}.

% state how to effectively align the intermediate features and the resulting architecture. It’s worth mentioning that our approach can also be generalized to other frame-wise video representation models (\eg, NeRV \cite{chen2021nerv}, E-NeRV \cite{li2022nerv}), with more details provided in appendix. 

\subsection{Overview}
\label{subsec:overview}
Let $\mathcal{X}=\{\boldsymbol{x}_1, \cdots, \boldsymbol{x}_T\}$ denote a video sequence, where $\boldsymbol{x}_t \in \mathbb{R}^{H \times W \times 3}$ is the frame at timestamp $t$ with height $H$ and width $W$. Following HNeRV \cite{chen2023hnerv}, we use ConvNeXt blocks \cite{liu2022convnet} to build a video-specific encoder $E$ that maps each individual frame $\boldsymbol{x}_t$ to a compact embedding $\boldsymbol{y}_t \in \mathbb{R}^{h \times w \times d}$ with $d$ representing the embedding dimension. Take a 1080$\times$1920 video as an example, we set $h=\frac{H}{120}$ and $w=\frac{W}{120}$. 
Note that the identity information in the original HNeRV is confined to the input embedding $\boldsymbol{y}_t$. We instead introduce a frame reconstruction network $F$ conditioned on the temporal embedding $\boldsymbol{z}_t$ to effectively align the intermediate features with identity information. Specifically, the frame index $t$, normalized to (0, 1], is initially mapped to a high-dimensional space using a regular frequency positional encoding function ${\rm PE}(\cdot)$ \cite{mildenhall2021nerf}, and then processed through a small MLP network $M$ to produce the temporal embedding $\boldsymbol{z}_t$. 
As depicted in Fig.~\ref{fig:arch}, when the input embedding $\boldsymbol{y}_t$ passes through the proposed SNeRV block, the size of the embedding usually increases step by step. Meanwhile, the temporal embedding $\boldsymbol{z}_t$ modulates the intermediate features in the proposed TAT residual block. In the end, a header layer transforms the output features of the last stage into the reconstructed frame $\boldsymbol{\hat{x}}_t$. Formally, the overall representation procedure is formulated as
\begin{equation}
\begin{aligned}
\label{equation:compression_procedure}
    \boldsymbol{y}_t = E(\boldsymbol{x}_t; \boldsymbol{\phi}), \boldsymbol{z}_t = M({\rm PE}(t); \boldsymbol{\psi}), \boldsymbol{\hat{x}}_t = F(\boldsymbol{y}_t, \boldsymbol{z}_t; \boldsymbol{\theta}) 
\end{aligned}
\end{equation}
% $$
% \boldsymbol{y}_t = E(\boldsymbol{x}_t; \boldsymbol{\phi}), \boldsymbol{z}_t = M({\rm PE}(t); \boldsymbol{\psi}), \boldsymbol{\hat{x}}_t = F(\boldsymbol{y}_t, \boldsymbol{z}_t; \boldsymbol{\theta})
% $$
where $\boldsymbol{\phi}$, $\boldsymbol{\psi}$, and $\boldsymbol{\theta}$ are the learnable parameters of the video-specific encoder, the temporal embedding generator, and the frame reconstruction network, respectively. The positional encoding function ${\rm PE}(t)$ is defined as $(\sin(b^0\pi t), \cos(b^0\pi t), \cdots, \sin(b^{l-1}\pi t), \cos(b^{l-1}\pi t))$ with hyperparameters $b$ and $l$.

\begin{figure}[t]
    \centering
    \includegraphics[scale=0.8]{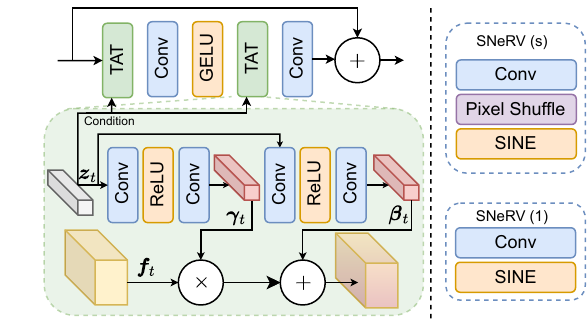}
    \caption{(Left) Illustration of the temporal-aware affine transform layer and residual block. The TAT layer takes the temporal embeddings $\boldsymbol{z}_t$ to produce channel-wise scaling and shifting parameters $\boldsymbol{\gamma}_t$ and $\boldsymbol{\beta}_t$. As a result, the affine transformation is performed to the intermediate features of the previous layer. (Right) The architecture of the sinusoidal NeRV-like block. When the stride $s$ of the convolutional layer is larger than 1, it includes a pixelshuffle layer.  
}
    \label{fig:tat}
    \vspace{-0.5cm}
\end{figure}

\subsection{Temporal-aware Conditional Decoder} 
\label{subsec:cond_decoder}

\noindent \textbf{Temporal-aware Affine Transform}.
Given the intermediate features $\boldsymbol{f}_t$ and temporal embedding $\boldsymbol{z}_t$, previous works \cite{li2022nerv, bai2023ps, gomes2023video} first adopt a small MLP network to learn the channel-wise mean $\boldsymbol{\mu}_t$ and variance $\boldsymbol{\sigma}_t$ of the target frame, and then use the AdaIN \cite{huang2017arbitrary} module to change the distribution of intermediate features:
\begin{equation}
\begin{aligned}
\label{equation:compression_procedure}
    (\boldsymbol{\mu}_t, \boldsymbol{\sigma}_t) &= {\rm MLP}(\boldsymbol{z}_t), \\
    {\rm AdaIN}(\boldsymbol{f}_t,\boldsymbol{\mu}_t, \boldsymbol{\sigma}_t)& = \boldsymbol{\sigma}_t (\frac{\boldsymbol{f}_t- \boldsymbol{\mu}(\boldsymbol{f}_t)}{\boldsymbol{\sigma}(\boldsymbol{f}_t)})+\boldsymbol{\mu}_t, 
\end{aligned}
\end{equation}
where $\boldsymbol{\mu}(\boldsymbol{f}_t)$ and $\boldsymbol{\sigma}(\boldsymbol{f}_t)$ are computed across spatial locations. However, the AdaIN module couples normalization and conditional affine transformation. The normalization operation typically serves to prevent the overfitting in neural networks, which conflicts with video INRs that utilize the overfitting to represent the data. 

To overcome this limitation, we present a temporal affine transform (TAT) layer without normalization and its associated residual block to unleash the potential of feature alignment. Fig.~\ref{fig:tat} (left) illustrates the details of our TAT residual block, which is inspired by the network design of \cite{wang2018recovering}. Based on the external temporal embedding $\boldsymbol{z}_t$, the TAT layer learns to generate a set of channel-wise affine parameters $(\boldsymbol{\gamma}_t, \boldsymbol{\beta}_t)$ for the intermediate features $\boldsymbol{f}_t$. Within this layer, the feature transformation is expressed as:  
\begin{equation}
\begin{aligned}
    {\rm TAT}(\boldsymbol{f}_t|\boldsymbol{\gamma}_t, \boldsymbol{\beta}_t) = \boldsymbol{\gamma}_t \boldsymbol{f}_t + \boldsymbol{\beta}_t,
\end{aligned}
\end{equation}
% By inserting the TAT residual block into existing representation models, aligned intermediate features can effectively improve the models' overfitting ability. 
By inserting the TAT residual block into existing video INRs, these aligned intermediate features can significantly enhance the models' overfitting ability. 

\begin{figure}[t]
  \subfloat[GELU activation]
  {\includegraphics[scale=0.19]{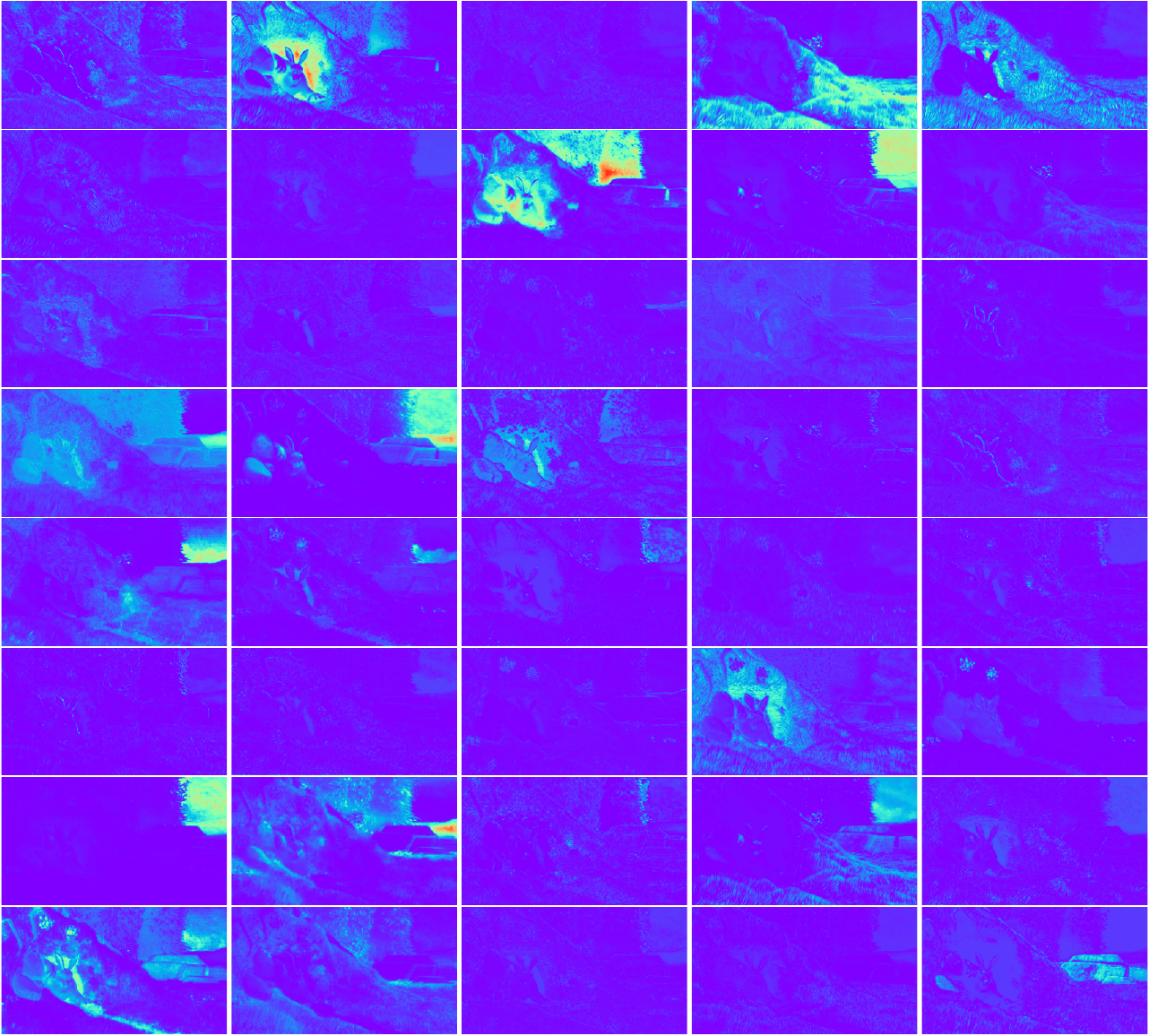}}
  \subfloat[SINE activation]
  {\includegraphics[scale=0.19]{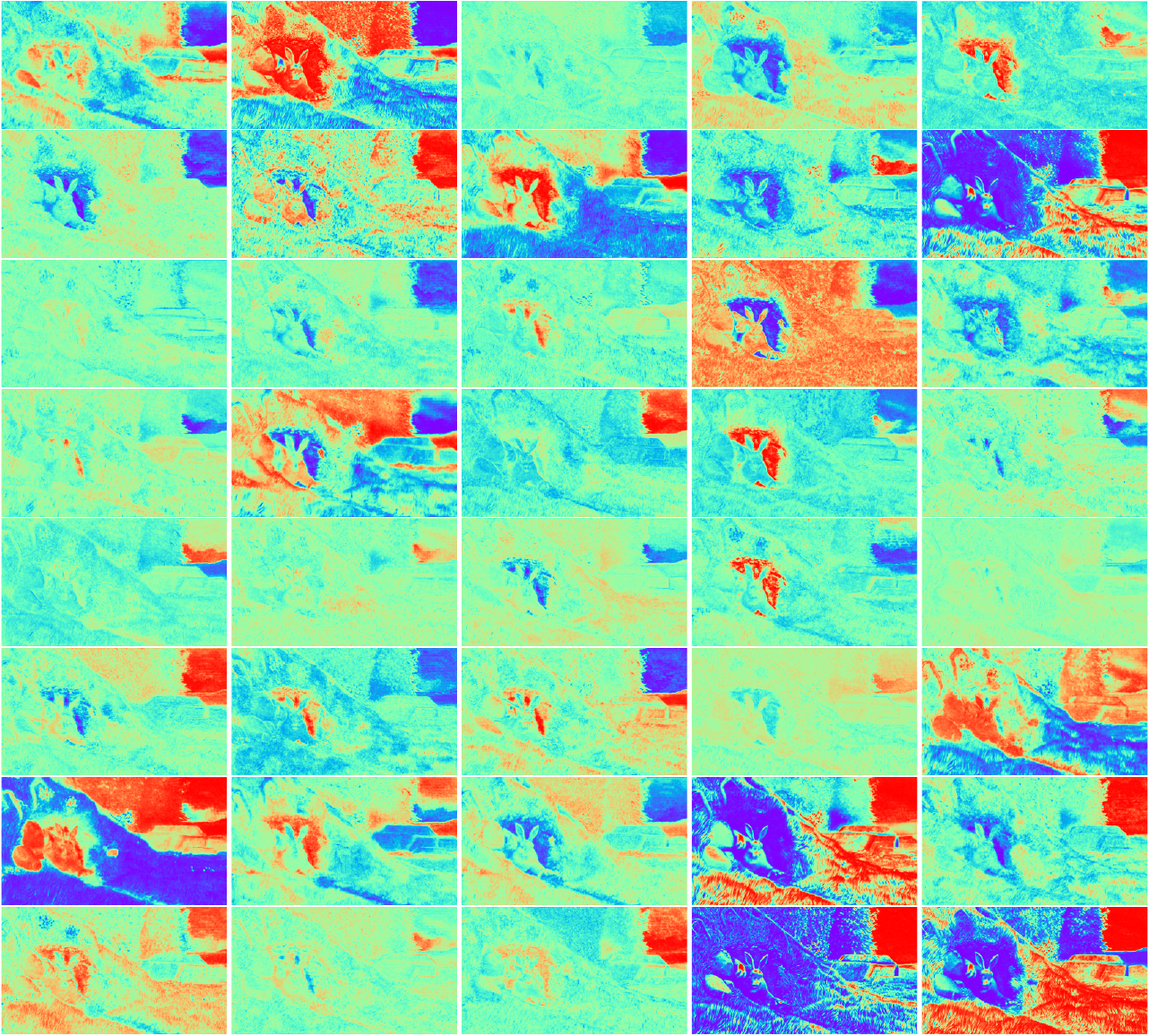}}
   \vspace{-0.3cm}
  \caption{Visual comparisons of intermediate features from different activation functions in the HNeRV-Boost model. We select the first 40 channel features from the last NeRV-like block on the first frame generation of the Bunny video.}
  \label{fig:feature_maps}
  \vspace{-0.5cm}
\end{figure}

\noindent \textbf{Sinusoidal NeRV-like Block}. Previous upsampling blocks \cite{chen2021nerv, li2022nerv, chen2022cnerv, lee2023ffnerv} commonly use GELU as their default activation function. However, our analysis of feature maps, as visualized in Fig.~\ref{fig:feature_maps}, reveals a limitation: the GELU layer tends to activate only a limited number of feature maps, whereas those activated by the SINE layer are more diverse and focus on different regions. This motivates us to introduce the sinusoidal NeRV-like (SNeRV) block. As shown in Fig.~\ref{fig:tat} (right), our SNeRV block has two types, where the one with a pixelshuffle layer serves for upsampling features.

Besides, we notice that the HNeRV blocks in the final three stages use a 5$\times$5 kernel size, resulting in about 2.7$\times$ more parameters than a 3$\times$3 kernel. However, reducing the kernel size directly adversely affects the regression performance. To resolve this difficulty, we substitute a single HNeRV block with a 5$\times$5 kernel for two SNeRV blocks with a 3$\times$3 kernel, setting the stride of the second SNeRV block to 1. This alteration enables us to maintain a similar level of video reconstruction quality with fewer parameters.

% \noindent \textbf{Compact SNeRV Block}. Previous upsampling blocks \cite{chen2021nerv, li2022nerv, chen2022cnerv, lee2023ffnerv} adopt GELU as their activation function by default. Fig.~\ref{fig:feature_maps} visualizes the feature maps after different activation functions in the HNeRV. We observe that only a few feature maps after GELU layer can be activated, while the feature maps after SINE layer appear quite diverse and focus on different regions. Thus, we design the SNeRV block that consists of three layers: convolution, pixelshuffle, and SINE. 

% In addition, the HNeRV blocks in the last three stages set the kernel size as 5, which yields approximately 2.7$\times$ the parameters compared to the use of 3$\times$3 kernel. However, simply reducing the kernel size greatly degrades the regression performance. Therefore, we replace an HNeRV block with 5$\times$5 kernel with two SNeRV blocks with 3$\times$3 kernel, where the stride of the second SNeRV block is set as 1. In this way, we achieve similar video reconstruction quality with fewer parameters. 

\noindent \textbf{Loss Function}. The objective of video INRs is to reduce the distortion between the original frame $\boldsymbol{x}_t$ and reconstructed frame $\boldsymbol{\hat{x}}_t$. Although the L1 loss is adept at preserving brightness as well as color, it falls short in maintaining high-frequency details. Therefore, we integrate a combination of the MS-SSIM and frequency domain losses into the L1 loss, ensuring a more comprehensive capture of high-frequency regions. For the frequency domain loss, we apply the fast Fourier transform (FFT) to both $\boldsymbol{x}_t$ and $\boldsymbol{\hat{x}}_t$, and then compute their L1 loss. The complete distortion loss function is formulated as follows:
\begin{equation}
  \begin{aligned}
  \mathcal{L}_d &= \mathcal{L}_1({\rm FFT}(\boldsymbol{x}_t), {\rm FFT}(\boldsymbol{\hat{x}}_t)) + \lambda \alpha \mathcal{L}_1(\boldsymbol{x}_t, \boldsymbol{\hat{x}}_t) \\
  &+\lambda (1-\alpha)(1- \mathcal{L}_{\rm MS\text{-}SSIM}(\boldsymbol{x}_t, \boldsymbol{\hat{x}}_t)) \label{eq:loss_func}
  \end{aligned}
\end{equation}
Here, $\lambda$ and $\alpha$ are hyperparameters used to balance the weight of each loss component.

\subsection{Consistent Entropy Minimization} %\subsection{Network-free Gaussian Entropy Model}
\label{subsec:compression}

After overfitting the video, we propose a consistent entropy minimization technique to refine the compression pipeline in \cite{gomes2023video, maiya2023nirvana}. As indicated in \tableautorefname~\ref{table:compression_comparision}, Gomes \textit{et al.} \cite{gomes2023video} and Maiya \textit{et al.} \cite{maiya2023nirvana} primarily concentrate on the quantization of model weights, overlooking the significance of embedding compression in improving the RD performance for hybrid-based video INRs. Furthermore, these methods use different entropy models in training and inference stages. Specifically, a small neural network is employed as a surrogate entropy model during training to estimate the bitrate, while in the inference phase, it is replaced with either CABAC or an arithmetic coder using a fixed statistical frequency table. This strategy aims to minimize transmission overhead from numerous small proxy networks. However, the discrepancy between the estimated bitrate by the surrogate model and the actual bitrate may mislead network optimization, resulting in sub-optimal coding efficiency. To overcome these shortcomings, our study introduces two key modifications to the entropy minimization pipeline: (i) applying a symmetric/asymmetric quantization scheme to model weights/embeddings, and (ii) introducing a network-free Gaussian entropy model with tiny metadata overhead to ensure the consistency during training and inference. %, which entails tiny metadata transmission overhead, ensures the consistency during training and inference.

\noindent \textbf{Quantization}. Since the model weights are empirically observed to be distributed symmetrically around zero \cite{esser2019learned, bhalgat2020lsq+}, a symmetric scalar quantization scheme with a trainable scale parameter $\varsigma$ is used for weights:
\begin{equation}
  \begin{aligned}
    Q(x)= \lfloor \frac{x}{\varsigma} \rceil, \; Q^{-1}(x)= x \times \varsigma
  \end{aligned}
\end{equation}
Conversely, for embeddings with skewed distributions, we adopt an asymmetric activation quantization scheme where both the scale parameter $\varsigma$ and the offset parameter $\eta$ are learned during training:
\begin{equation}
  \begin{aligned}
    Q(x)= \lfloor \frac{x-\eta}{\varsigma} \rceil, \; Q^{-1}(x)= x \times \varsigma + \eta
  \end{aligned}
\end{equation}
Given the non-differentiable nature of the quantization operation during training, we leverages a mixed quantizer outlined in \cite{gomes2023video, maiya2023nirvana} to allow end-to-end optimization. Specifically, the rounding operation is substituted with uniform noise $\mathcal{U}(-\frac{1}{2}, \frac{1}{2})$ for entropy calculation, while the straight-through estimator (STE) is applied for distortion calculation when computing the gradient for the rounding operation.

\noindent \textbf{Network-free Gaussian Entropy Model}. We model the probability of the quantized embedding $\boldsymbol{\hat{y}}_t$ with a Gaussian distribution:
\begin{equation}
\begin{aligned}
p(\boldsymbol{\hat{y}}_t) = \prod_{i} \big(\mathcal{N}(\mu_{\boldsymbol{y}_t}, \sigma_{\boldsymbol{y}_t}^{2})*\mathcal{U}(-\frac{1}{2}, \frac{1}{2})\big)(\hat{y}_t^i)
% p(\boldsymbol{\hat{y}}_t) \sim \mathcal{N}(\mu_{\boldsymbol{y}_t}, \sigma_{\boldsymbol{y}_t}^{2})
\end{aligned}
\end{equation}
where $\mu_{\boldsymbol{y}_t}$ and $\sigma_{\boldsymbol{y}_t}^{2}$ are the mean and variance of the embedding $\boldsymbol{y}_t$. $*$ denotes convolution. Similarly, we independently compute the mean and variance of each weight as the entropy model parameters for weight compression. This novel entropy model brings two main advantages. On the one hand, when compared with the neural entropy model of \cite{gomes2023video, maiya2023nirvana} that only captures the relationship between elements of each kernel, our model can capture the global relationships within all elements in the weight, which facilitates the accurate estimation of the probability distribution. On the other hand, thanks to only transmitting two scalar values for each weight/embedding, our entropy model can get rid of the surrogate position and directly provide the probability distribution for arithmetic coding in the inference stage. 

\noindent \textbf{Optimization Objective}. The goal of INR compression is to achieve high reconstruction quality with minimal bitrate consumption. To this end, we incorporate an entropy regularization term $\mathcal{L}_r$ to encourage smaller compressed models. In order to control the whole compression ratio, we introduce $R_{target}$ into the regularization term. Once $R_{target}$ is satisfied, $\mathcal{L}_r$ diminishes to zero:
\begin{equation}
  \begin{aligned}
    \mathcal{L} = \mathcal{L}_d + \kappa \mathcal{L}_r = \mathcal{L}_d + \kappa {\rm ReLU}(R-R_{target}) \label{eq:cem_loss}
  \end{aligned}
\end{equation}
Here, $\kappa$ is a hyperparameter balancing the compression rate and distortion. $R$ is calculated as $\frac{\sum_{t=1}^{T} R(\boldsymbol{\hat{y}}_{t}) +  R(\boldsymbol{\hat{\theta}})+R(\boldsymbol{\hat{\psi}})} {T\times H \times W}$, where $R(\boldsymbol{\hat{y}}_{t})$ denotes the estimated bitrate of the quantized embedding $\boldsymbol{\hat{y}}_{t}$ and $R(\boldsymbol{\hat{\theta}})/R(\boldsymbol{\hat{\psi}})$ represents the estimated bitrate of the quantized weights $\boldsymbol{\hat{\theta}}/\boldsymbol{\hat{\psi}}$. For $R_{target}$, we define it as $B_{avg}\frac{\sum_{t=1}^{T}{\rm Numel}(\boldsymbol{y}_t)+{\rm Numel}(\boldsymbol{\theta})+{\rm Numel}(\boldsymbol{\psi})}{T\times H \times W}$, in which $B_{avg}$ indicates the average bit-width of the compressed INR and ${\rm Numel}(\cdot)$ means the amount of parameters. 

\begin{table}
\setlength{\tabcolsep}{2pt}
\scriptsize
  \caption{Comparisons between different entropy minimization techniques in INR compression.}
  \vspace{-0.25cm}
  \label{table:compression_comparision}
  \centering
  \begin{tabular}{c|cc|cc}
    \hline
    \multirow{2}*{Method} &  \multicolumn{2}{c|}{Quantization} & \multicolumn{2}{c}{Entropy Model}  \\
    & Weight & Embedding & Training & Inference \\
    \hline
    Gomes \textit{et al.}\cite{gomes2023video} & Asymmetric & - & Neural network & CABAC \\
    Maiya \textit{et al.}\cite{maiya2023nirvana} & Symmetric & - & Neural network & Fixed frequency table \\
    \hline
    CEM (ours)	& Symmetric & Asymmetric & \multicolumn{2}{c}{Network-free Gaussian entropy model} \\
    \hline
  \end{tabular}
  \vspace{-0.5cm}
\end{table}

\section{Experiments}
\label{sec:experiments}

\subsection{Experimental Setup}

\textbf{Dataset.} We evaluate the effectiveness of our framework on multiple benchmarks with various kinds of video contents, including Bunny \cite{bunny2010} (720$\times$1280 with 132 frames), UVG \cite{mercat2020uvg} (7 videos at 1080$\times$1920 with length of 600 or 300), and DAVIS validation \cite{perazzi2016benchmark} (20 videos at 1080$\times$1920). 

\noindent \textbf{Evaluation Metrics.} Two popular image quality assessment metrics, namely, PSNR and MS-SSIM, are used to evaluate the distortion between the reconstructed and original frames. We use bits per pixel (bpp) to measure the bitrate of video compression.

\noindent \textbf{Implementation Details.}
We select three typical INR methods (\ie, NeRV \cite{chen2021nerv}, E-NeRV \cite{li2022nerv}, and HNeRV \cite{chen2023hnerv}) as our baselines. Then we enhance them with our proposed framework. To fit 720p and 1080p videos, we set the stride list as (5,2,2,2,2) and (5,3,2,2,2), respectively. We use $b=1.25$ and $l=80$ as our default setting in the position encoding. For the distortion loss in \equationautorefname~\ref{eq:loss_func}, $\lambda$ and $\alpha$ are set as 60 and 0.7, receptively. During overfitting, we set the batch size as 1 and adopt Adan \cite{xie2022adan} as the optimizer with cosine learning rate decay \cite{chen2023hnerv}, in which the number of warm-up epochs is 10\% of total fitting epochs. The learning rates for the boosted E-NeRV and NeRV/HNeRV are $1.5e^{-3}$ and $3e^{-3}$, respectively. Baseline models are implemented using open-source codes, and experiments are conducted on one NVIDIA GTX 1080Ti GPU using PyTorch, with 3M model size and 300 epochs unless otherwise denoted. For more details about  different tasks, please refer to the supplementary material.

% After overfitting the model, we fine-tune the model to perform entropy minimization for 100 epochs. We empirically found that setting $B_{avg}$ as 4 bits can achieve a better RD trade-off. For more details about different tasks, please refer to the supplementary material.

% \subsection{Video Regression}
% \begin{table}
% \scriptsize
%   \caption{PSNR on Bunny with different \textbf{sizes}.}
%   \label{table:size_bunny}
%   \centering
%   \begin{tabular}{c|cccc}
%     \hline
%     Size & 0.75M & 1.5M & 3M & Avg. \\
%     \hline
%     NeRV  & 26.12 & 28.54 & 31.84 & 28.83  \\
%     NeRV-Boost & \textbf{30.25} & \textbf{33.71} & \textbf{37.25} & \textbf{33.74} \\
%     \hline
%     E-NeRV & 29.39 & 33.44 & 37.32 & 33.38 \\
%     E-NeRV-Boost & \textbf{32.61} & \textbf{37.19} & \textbf{40.07} & \textbf{36.62} \\
%     \hline
%     HNeRV & 30.55 & 35.13 & 38.15 & 34.61  \\
%     HNeRV-Boost & \textbf{35.09} & \textbf{38.52} & \textbf{41.09} & \textbf{38.23} \\
%     \hline
%   \end{tabular}
%   \vspace{-0.25cm}
% \end{table}

\subsection{Video Regression}
\begin{table}
\footnotesize
  \caption{Average PSNR on UVG with various model sizes.}
    \vspace{-0.25cm}
  \label{table:uvg_size}
  \centering
  \begin{tabular}{c|ccccc}
    \hline
    Size & 3M & 5M & 10M & 15M & Avg. \\
    \hline
    NeRV \cite{chen2021nerv}	    &31.11	&32.44	&34.20	&35.18 &33.23\\
    NeRV-Boost	&\textbf{32.76}&\textbf{33.76}&\textbf{35.28}&\textbf{35.94}&\textbf{34.43} \\
    \hline
    E-NeRV \cite{li2022nerv}	    &31.51	&33.99	&34.42	&35.32& 33.81 \\
    E-NeRV-Boost	&\textbf{33.40}&\textbf{34.31}&\textbf{35.58}&\textbf{36.27}&\textbf{34.89}\\
    \hline
    HNeRV \cite{chen2023hnerv}	    &32.47	&33.40	& 34.70	& 35.10	& 33.92  \\
    HNeRV-Boost	&\textbf{33.89}&\textbf{35.06}&\textbf{36.49}&\textbf{37.29}&\textbf{35.68}\\
    \hline
  \end{tabular}
  \vspace{-0.25cm}
\end{table}

\begin{table}
\setlength{\tabcolsep}{2pt}
\scriptsize
  \caption{Detailed PSNR of each video on the UVG dataset with 3M model size.}
    \vspace{-0.25cm}
  \label{table:uvg_details}
  \centering
  \begin{tabular}{c|cccccccc}
    \hline
    Video &  Beauty & Bosph. & Honey. & Jockey & Ready. & Shake. & Yacht. & Avg. \\
    \hline
    NeRV \cite{chen2021nerv} & 33.14 &	32.74 &	37.18 &	30.99 &	23.97 &	33.06 &	26.72 &	31.11 \\
    NeRV-Boost & \textbf{33.55} & \textbf{34.51}&\textbf{39.04}&\textbf{32.82} &\textbf{26.08} & \textbf{34.54} & \textbf{28.76} & \textbf{32.76} \\
    \hline
    E-NeRV \cite{li2022nerv} & 33.29 & 33.87 & 38.88 & 28.73 & 23.98 & 34.45 &  27.38 & 31.51 \\
    E-NeRV-Boost & \textbf{33.75} & \textbf{35.62} & \textbf{39.61} & \textbf{32.39} & \textbf{27.75} & \textbf{35.48} & \textbf{29.23} & \textbf{33.40} \\
    \hline
    HNeRV \cite{chen2023hnerv} &33.36 &	33.62 &	39.17 &	32.31 &	25.60 &	34.90 &	28.33 &	32.47 \\
    HNeRV-Boost & \textbf{33.80} & \textbf{36.11} & \textbf{39.65} & \textbf{34.28} & \textbf{28.19} & \textbf{35.88} & \textbf{29.33} & \textbf{33.89} \\
    \hline
  \end{tabular}
  \vspace{-0.25cm}
\end{table}

\begin{table}
\footnotesize
  \caption{PSNR on the Bosphorus video with different epochs.}
    \vspace{-0.25cm}
  \label{table:bosph_epochs}
  \centering
  \begin{tabular}{c|ccccc}
    \hline
    Epoch & 300 & 600 & 1200 & 1800 & 2400 \\
    \hline
    NeRV \cite{chen2021nerv} & 32.74 & 33.00 & 33.20 & 33.27 &33.32 \\
    NeRV-Boost & \textbf{34.51} & \textbf{34.73} & \textbf{34.89} & \textbf{34.97} & \textbf{35.02}  \\
    \hline
    E-NeRV \cite{li2022nerv} & 33.87 & 34.19 & 34.40 & 34.50 & 34.56 \\
    E-NeRV-Boost& \textbf{35.62} & \textbf{35.92} & \textbf{36.16} & \textbf{36.27} & \textbf{36.32} \\
    \hline
    HNeRV \cite{chen2023hnerv} & 33.62 & 34.15 & 34.35 & 34.41 & 34.46 \\
    HNeRV-Boost& \textbf{36.11} & \textbf{36.33} & \textbf{36.52} & \textbf{36.59} & \textbf{36.64} \\
    \hline
  \end{tabular}
  \vspace{-0.45cm}
\end{table}

\tableautorefname~\ref{table:uvg_size} show the regression performance of various methods on the UVG dataset at different scales. It is evident that our boosted methods achieve superior reconstruction quality over the corresponding baselines. As detailed in \tableautorefname~\ref{table:uvg_details}, the improvements are consistent across all test videos in the UVG dataset. For instance, compared to the NeRV, E-NeRV, and HNeRV baselines on the ReadySetGo video, our boosted versions exhibit considerable improvements of about 2.11dB, 3.77dB, and 2.59dB, respectively. In \tableautorefname~\ref{table:bosph_epochs}, we offer a comparison of regression performance between the boosted versions and baselines on the Bosphorus video across different fitting epochs. Notably, our boosted versions at 300 epochs outperform the baselines at 2400 epochs.
Furthermore, as shown in Fig.~\ref{fig:buuny_regression} (top), our boosted versions at minimal training time surpass the baselines at their maximum training time by a significant margin on the Bunny video, which indicates the superiority of our method in accelerating convergence speed and improving the representation capabilities.

\begin{figure}[t]
 \centering
  \subfloat
  {\includegraphics[scale=0.3]{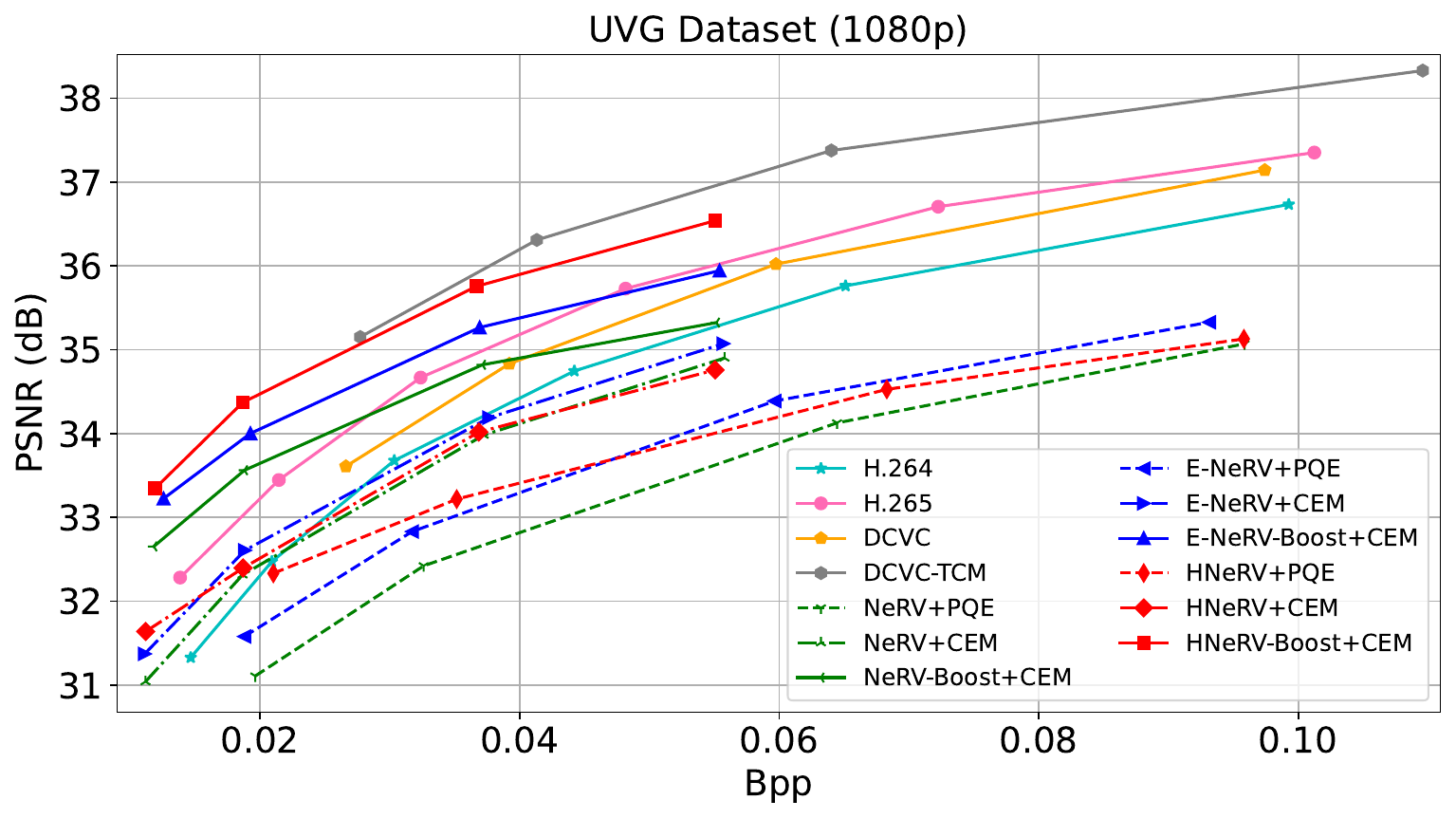}}\\
  \subfloat
  {\includegraphics[scale=0.3]{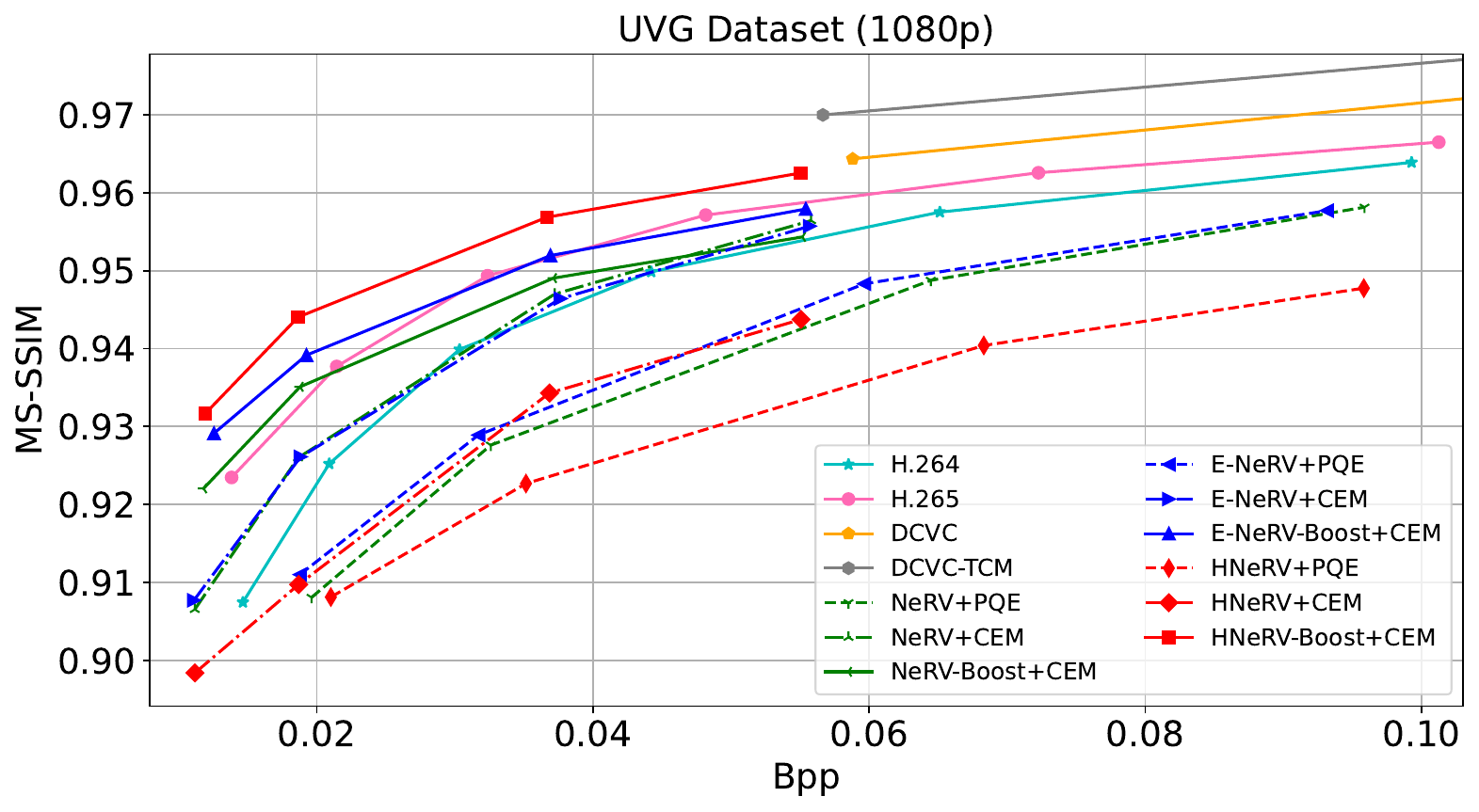}}
   \vspace{-0.3cm}
  \caption{Rate-distortion curves of our boosted approaches and different baselines on the UVG dataset in PSNR and MS-SSIM. PQE denotes the three-step compression pipeline of NeRV.}
  \label{fig:rd_performance}
  \vspace{-0.2cm}
\end{figure}

\begin{table}
\setlength{\tabcolsep}{2pt}
\footnotesize
  \caption{Complexity comparison at resolution 1920$\times$1080. The decoding latency is evaluated by an NVIDIA V100 GPU.}
    \vspace{-0.25cm}
  \label{table:complexity}
  \centering
  \begin{tabular}{l|ccc}
    \hline
   Method   & Params $\downarrow$  & Decoding time $\downarrow$& FPS $\uparrow$ \\
   \hline
   DCVC \cite{li2021deep} & 35.2M &  35590ms  &0.028 \\
   DCVC-TCM \cite{sheng2022temporal}& 40.9M & 470ms & 2.12 \\
   \hline
   NeRV \cite{chen2021nerv} & 3.04M& 7ms& 135.64\\
   NeRV-Boost  &3.06M & 23ms& 43.54\\
   \hline
   E-NeRV \cite{li2022nerv} & 3.01M& 18ms& 54.75\\
   E-NeRV-Boost  &3.03M & 53ms & 18.74\\
   \hline
   HNeRV \cite{chen2023hnerv}& 3.05M& 41ms& 24.22\\
   HNeRV-Boost & 3.06M& 76ms & 13.15 \\
    \hline
  \end{tabular}
  \vspace{-0.35cm}
\end{table}

\begin{table*}[t]
\setlength{\tabcolsep}{2pt}
\scriptsize
  \caption{Video inpainting results on the DAVIS validation dataset in PSNR. Mask-S and Mask-C refers to disperse and central mask scenarios, respectively.}
    \vspace{-0.25cm}
  \label{table:davis_details}
  \centering
  \begin{tabular}{l|cccccc|cccccc}
    \hline
    \multirow{2}*{Video} &  \multicolumn{6}{c|}{Mask-S} & \multicolumn{6}{c}{Mask-C}  \\
    &  NeRV & NeRV-Boost  & E-NeRV & E-NeRV-Boost & HNeRV & HNeRV-Boost  & NeRV & NeRV-Boost  & E-NeRV & E-NeRV-Boost & HNeRV & HNeRV-Boost  \\
    \hline
Blackswan	        &27.06	&\textbf{30.46} & 29.53 & \textbf{31.34} & 30.20 & \textbf{34.10} & 24.11 & \textbf{26.89} & 26.38 & \textbf{27.88} & 26.45 & \textbf{29.18} \\
Bmx-trees	        &26.77	&\textbf{30.16} & 27.75 & \textbf{30.86} & 29.05 & \textbf{32.99} & 22.43 & \textbf{25.14} & 23.79 & \textbf{26.66} & 22.28 & \textbf{22.28} \\
Breakdance	        &25.48	&\textbf{28.46} & 26.97 & \textbf{30.57} & 26.34 & \textbf{33.10} & 20.16 & \textbf{22.28} & 22.15 & \textbf{22.15} & 20.23 & \textbf{20.24} \\
Camel	            &23.70	&\textbf{26.09} & 25.70 & \textbf{27.56} & 26.13 & \textbf{31.08} & 21.21 & \textbf{23.16} & 22.62 & \textbf{23.55} & 17.74 & \textbf{19.81} \\
Car-roundabout	    &23.92	&\textbf{28.25} & 26.32 & \textbf{29.43} & 28.64 & \textbf{31.90}  & 21.24 & \textbf{23.53} & 22.73 & \textbf{24.51} & 21.71 & \textbf{22.36} \\
Car-shadow	        &26.58	&\textbf{32.40} & 30.63 & \textbf{33.00}    & 31.01 & \textbf{35.85} & 23.07 &	\textbf{24.13} & 23.21 & \textbf{24.10}  & 21.05 & \textbf{23.65} \\
Cows	            &22.17	&\textbf{24.77} & 23.92 & \textbf{26.41} & 24.68 & \textbf{28.30}  & 20.48 &	\textbf{22.39} & 21.88 & \textbf{23.13} & 21.82 & \textbf{24.14} \\
Dance-twirl	        &25.29	&\textbf{28.49} & 27.42 & \textbf{29.38} & 28.74 & \textbf{30.79} & 21.17 &	\textbf{23.14} & 22.40  & \textbf{23.34} & 21.06 & \textbf{21.77} \\
Dog	                &29.29	&\textbf{31.97}	& 31.72 & \textbf{32.79} & 28.80  & \textbf{33.87} & 25.37 & \textbf{27.02} & 27.07 & \textbf{28.25} & 24.16 & \textbf{24.66} \\
Drift-chicane       &34.09	&\textbf{39.94} & 39.26 & \textbf{41.60}  & 38.52 & \textbf{43.32} & 27.52 &	\textbf{28.01} & 29.81 & \textbf{31.52} & 23.40  & \textbf{27.44} \\
Drift-straight      &26.78	&\textbf{32.26} & 29.53 & \textbf{33.19} & 30.81 & \textbf{36.16} & 22.76 &	\textbf{26.00} & 24.69 & \textbf{27.12} & 18.88 & \textbf{21.49} \\
Goat	            &24.04	&\textbf{26.30} & 25.34 & \textbf{27.21}  & 26.91 & \textbf{30.59} & 22.03 &	\textbf{23.90}  & 23.43 & \textbf{24.56} & 23.06 & \textbf{25.10} \\
Horsejump-high      &25.74	&\textbf{30.39} & 29.27 & \textbf{31.26} & 29.31 & \textbf{30.86} & 21.54 &	\textbf{23.46} & 23.06 & \textbf{23.93} & 20.72 & \textbf{23.16} \\
Kite-surf	        &29.34	&\textbf{34.18} & 32.87 & \textbf{35.16} & 33.49 & \textbf{37.08} & 23.92 &	\textbf{27.22} & 26.71 & \textbf{28.87} & 24.73 & \textbf{27.49} \\
Libby	            &29.81	&\textbf{34.24} & 31.39 & \textbf{34.95} & 28.66 & \textbf{37.35} & 25.71 &	\textbf{28.14} & 26.91 & \textbf{28.95} & 23.39 & \textbf{26.96} \\
Motocross-jump      &29.82	&\textbf{37.36} & 34.15 & \textbf{36.92} & 28.27 & \textbf{36.42} & 26.19 &	\textbf{29.65} & 28.75 & \textbf{29.30} & 22.36 & \textbf{26.25} \\
Paragliding-launch  &29.03	&\textbf{31.40} & 30.62 & \textbf{32.28}  & 30.99 & \textbf{33.64} & 25.95 &	\textbf{26.97} & 26.65 & \textbf{27.41} & 26.00	& \textbf{28.07} \\
Parkour	            &24.74	&\textbf{27.19} & 25.62 & \textbf{27.54} & 26.34 & \textbf{28.79} & 22.32 &	\textbf{24.48} & 22.99 & \textbf{24.43} & 19.06 & \textbf{20.55} \\
Scooter-black       &23.35	&\textbf{27.75} & 26.46 & \textbf{29.07} & 28.41 & \textbf{30.42} & 19.24 &	\textbf{21.77} & 20.99 & \textbf{22.14} & 18.94 & \textbf{19.86} \\
Soapbox	            &27.20	&\textbf{30.56} & 28.83 & \textbf{31.44} & 30.30  & \textbf{32.95} & 22.29 &	\textbf{25.00} & 23.82 & \textbf{25.51} & 17.98 & \textbf{19.20} \\
\hline
Average	            &26.71	&\textbf{30.63} & 29.17 & \textbf{31.60}  & 29.28 & \textbf{33.48} & 22.94 &	\textbf{25.11} & 24.50  & \textbf{25.87} & 21.75 & \textbf{23.68} \\
    \hline
  \end{tabular}
  \vspace{-0.35cm}
\end{table*}

\subsection{Video Compression}
To assess video compression performance, we follow \cite{chen2023hnerv} to train a model for each video, rather than encoding all videos together using a single network as in \cite{chen2021nerv, bai2023ps}. After model overfitting, we fine-tune these models using our entropy minimization method for 100 epochs with the initial learning rate $5e{-4}$ and cosine learning rate decay. We empirically choose $B_{avg}$ as 4 bits to optimize the RD trade-off. For the baselines, both the three-step compression pipeline of NeRV and our CEM technique are applied for comparison. Besides, we compare our boosted models with traditional codecs (H.264 \cite{wiegand2003overview}, H.265 \cite{sullivan2012overview}) and state-of-the-art (SOTA) learning-based codecs (DCVC \cite{li2021deep}, DCVC-TCM \cite{sheng2022temporal}), where H.264 and H.265 are tested using FFmpeg with the \textit{veryslow} preset and enabling B frames. 

Fig.~\ref{fig:rd_performance} presents the RD curves of these methods on the UVG dataset. Our boosted models offer remarkable improvements over their corresponding baselines, indicating the generalization of our framework. Notably, the original HNeRV shows limited robustness in compression, which constrains the efficacy of content-relevant embeddings, especially at higher bitrates. On the contrary, with our modifications, the boosted HNeRV consistently surpasses DCVC, H.265, H.264, and other INR methods across all bitrates in terms of PSNR, which underscores the ability of our framework in amplifying the advantages of the INR model itself. Additionally, baselines with CEM outperform those using the three-step compression, highlighting the effectiveness of our compression technique in enhancing RD performance. The complexity comparison of video decoding with two SOTA neural codecs is shown in \tableautorefname~\ref{table:complexity}.

\subsection{Video Inpainting}
In this section, we evaluate video inpainting on the DAVIS validation dataset using both disperse \cite{chen2023hnerv} and central masks \cite{zhao2023dnerv}. For the disperse mask scenario, each frame is overlaid with five uniformly distributed square masks of width 50. In the central mask scenario, a single rectangular mask, spanning one quarter of the frame's width and height, is centrally placed. Consistent with \cite{chen2023hnerv}, the distortion loss during training is calculated only for the non-masked pixels. During inference, the masked regions are reconstructed using the output from video INRs.

As shown in \tableautorefname~\ref{table:davis_details}, our boosted versions significantly improve the inpainting performance of the original baselines. Under the disperse mask case, we observe an average improvement of 3.92dB, 2.43dB, and 4.2dB for NeRV, E-NeRV, and HNeRV, respectively. Even in the more challenging central mask scenario, our enhanced versions still achieve improvements of 2.17dB, 1.37dB, and 1.93dB. 

\begin{table}
\setlength{\tabcolsep}{2pt}
\scriptsize
  \caption{Video interpolation results on the UVG dataset in PSNR.}
    \vspace{-0.25cm}
  \label{table:uvg_interpolation}
  \centering
  \begin{tabular}{c|cccccccc}
    \hline
    Video &  Beauty & Bosph. & Honey. & Jockey & Ready. & Shake. & Yacht. & Avg. \\
    \hline
    NeRV \cite{chen2021nerv}	    & \textbf{31.26} &	32.21 &	36.84 &	\textbf{22.24} & \textbf{20.05} &	32.09 &	26.09 &	28.68 \\
    NeRV-Boost	& 31.06 &	\textbf{34.28} &	\textbf{38.83} &	21.74 &	19.88 &	\textbf{32.58} &	\textbf{27.07} &	\textbf{29.35} \\
    \hline
    E-NeRV \cite{li2022nerv}	    & 31.25 &	33.36 &	38.62 &	\textbf{22.35} &	20.08 &	\textbf{32.82} &	26.74 &	29.32 \\
    E-NeRV-Boost	& \textbf{31.35} &	\textbf{35.01} &	\textbf{39.24} &	21.96 &	\textbf{20.45} &	32.75 &	\textbf{27.79} &	\textbf{29.79} \\
    \hline
    HNeRV \cite{chen2023hnerv}	    & 31.42 &	34.00 &	39.07 &	23.02 &	20.71 &	32.58 &	26.74 &	29.65 \\
    HNeRV-Boost	& \textbf{31.61} &	\textbf{36.16} &	\textbf{39.38} &	\textbf{23.14} &	\textbf{21.61} &	\textbf{32.94} &	\textbf{28.01} &	\textbf{30.41} \\
    \hline
  \end{tabular}
  \vspace{-0.35cm}
\end{table}

\subsection{Video Interpolation}
We evaluate the video interpolation performance of our boosted models on the UVG dataset. In this experiment, we use the odd-numbered frames of each video as the training set and the even-numbered frames as the test set. Following the approach in \cite{li2022nerv}, we adjust the frequency value $b$ to 1.05 for achieving better interpolation results while maintaining robust regression performance on the training set. The quantitative outcomes are detailed in \tableautorefname~\ref{table:uvg_interpolation}, These results indicate that our boosted models outperform the baselines in terms of overall interpolated quality.
%and the qualitative performance can be observed in Fig.~\ref{fig:visualizations}. These results indicate that our boosted models outperform the baselines in terms of overall interpolated quality.

% We conduct video interpolation experiments on the UVG dataset to evaluate the generalization of our augmented versions. The odd frames of each video is taken out as a training set and the remaining frames consist of a test set. According to \cite{li2022nerv}, we adjust the frequency value $b$ as 1.05 to achieve better interpolation while preserving the regression performance on the training set. The quantitative results are listed in \tableautorefname~\ref{table:uvg_interpolation} and the qualitative performance is shown in Fig.~\ref{fig:visualizations}. These results demonstrate our revised versions are superior to the corresponding baselines in terms of overall performance.

\subsection{Ablation Study}
\label{subsec:ablation}

\begin{table}
\setlength{\tabcolsep}{2pt}
\footnotesize
  \caption{Ablation studies for different boosting components on the Bunny video over 300 epochs, with results presented in PSNR.}
    \vspace{-0.25cm}
  \label{table:ablation}
  \centering
  \begin{tabular}{l|ccc}
    \hline
   Variant  & NeRV-Boost  & E-NeRV-Boost & HNeRV-Boost  \\
    \hline
    Ours	     &\textbf{37.25} &\textbf{40.07} &\textbf{41.09} \\
    \hline
    (V1) w/o TAT	&34.63	&35.75 &  39.12 \\
    (V2) w/ AdaIN	&35.59	&39.51 &38.03 \\
    (v3) w/ SAF     &34.28  &39.62 &40.93 \\
    \hline
    (V4) w/ GELU &34.85	&38.34 & 41.00 \\
    \hline
    (V5) w/ L2 &35.32	&38.55 & 40.19 \\
    (V6) w/ L1+SSIM &36.28	&39.34 & 41.00 \\
    (V7) w/ L1 &34.75	&37.76 & 40.37 \\
    (V8) w/ L1+MS-SSIM &36.12	&38.66 & 40.49 \\
    (V9) w/ L1+freq. &37.12	&39.60 & 41.08 \\
    (V10) w/ L1+SSIM+freq. & 36.99 & 40.05 & 41.01 \\
    \hline
  \end{tabular}
  \vspace{-0.25cm}
\end{table}

% \begin{figure}[t]
%     \centering
%     \includegraphics[scale=0.65]{figure/visual.pdf}
%      \vspace{-0.6cm}
%     \caption{Visualization comparison of disperse inpainting, central inpainting, and interpolation on the DAVIS validation and UVG datasets. The top row displays the ground truth, followed by our boosted results in (a, c, e) and the baseline results in (b, d, f). The red and blue numbers indicate the PSNR values for the respective colored patches.}
%     \label{fig:visualizations}
%     \vspace{-0.3cm}
% \end{figure}

\begin{table}
\setlength{\tabcolsep}{2pt}
\footnotesize
  \caption{Ablation studies for various upsampling blocks in the HNeRV-Boost framework on the Bunny video. GELU and SINE represent the activation function employed in different blocks. STD refers to the standard deviation of the model parameters' distribution, where a lower STD value signifies a more uniform distribution of model parameters.}
    \vspace{-0.25cm}
  \label{table:block_ablation}
  \centering
  \begin{tabular}{c|ccccc}
    \hline
   Block  & NeRV \cite{chen2021nerv} & E-NeRV \cite{li2022nerv}  & FFNeRV \cite{lee2023ffnerv} & HNeRV \cite{chen2023hnerv} & SNeRV  \\
   \hline
   GELU  & 39.61 & 39.26 & 39.33  & 40.77 & \textbf{41.00}\\
   SINE  & 40.35 & 39.99  & 40.06 & 40.93 & \textbf{41.09} \\
   \hline
    STD  & 0.225 & 0.208 & 0.176 & 0.047 & 0.045\\
    \hline
  \end{tabular}
  \vspace{-0.35cm}
\end{table}

\textbf{Feature Modulation}. We conduct a sets of ablation studies on Bunny to evaluate the effectiveness of our TAT module. In the first variant (V1), the TAT residual blocks are removed from our boosted INR models. As \tableautorefname~\ref{table:ablation} shows, this variant experiences an average drop of 2.97dB, implying the significance of using temporal embeddings to modulate intermediate features for accurate frame generation. In the second variant (V2), the TAT block is replaced with an AdaIN \cite{huang2017arbitrary} module, resulting in an average decrease in PSNR by 1.76dB and 1.19dB. It suggests that integrating normalization into the conditional affine transformation limits the overfitting capabilities of INR models to some degree. Furthermore, we compare the spatially-adaptive fusion (SAF) \cite{he2023towards} block. The results in variant V3 imply the superior temporal alignment capability of our TAT module.

% and a spatially-adaptive fusion (SAF) \cite{he2023towards} block, respectively. These replacements results in an average decrease in PSNR by 1.76dB and 1.19dB. It suggests that integrating normalization into the conditional affine transformation limits the overfitting capabilities of INR models to some degree.

% \textbf{Feature Modulation}. To show the superiority of our TAT residual block, two sets of ablation studies are conducted on Bunny. First, we remove the TAT residual block (variant V1) in our augmented INR models. As shown in \tableautorefname~\ref{table:ablation}, the variant V1 performs worse and drops by about 2.97dB on average, which demonstrates that using temporal embeddings to modulate intermediate features helps guide the model to generate the corresponding frame. We further report the results when the TAT residual block is directly replaced by AdaIN module (variant V2). It leads to a decrease of PSNR by 1.66dB, 0.56dB, 3.06dB compared with our approach, which indicates that incorporating the redundant normalization into conditional affine transform inhibits the overfitting potential of the INR model.

\begin{figure}[t]
    \centering
    \includegraphics[scale=0.4]{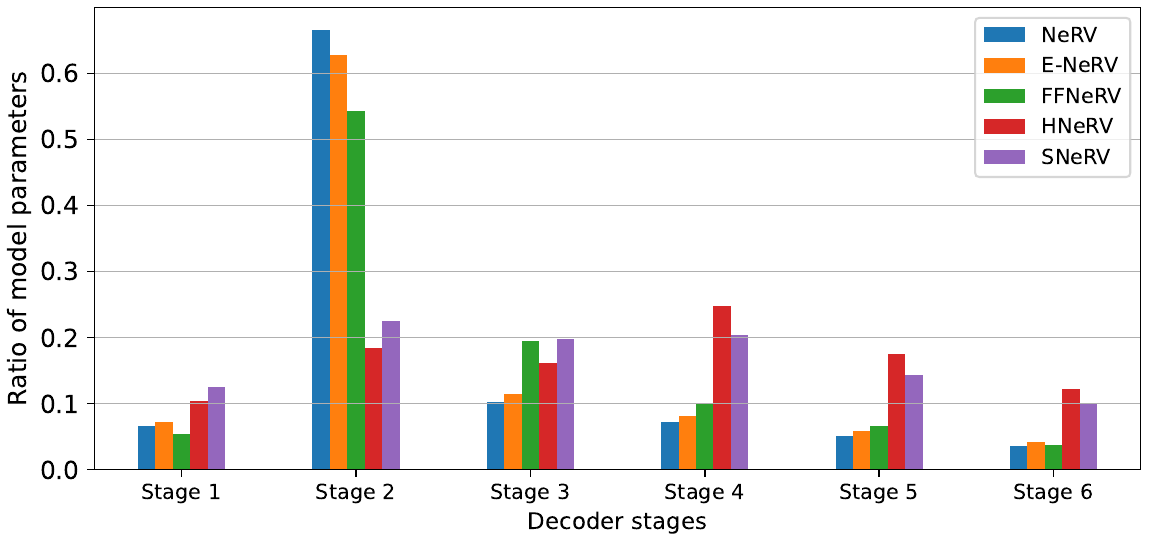}
    \vspace{-0.3cm}
    \caption{Distribution of model parameters across various decoder blocks in our HNeRV-Boost framework. See \tableautorefname~\ref{table:block_ablation} for PSNR results under these five configurations.}
    \label{fig:params}
    \vspace{-0.4cm}
\end{figure}

\noindent \textbf{NeRV-like Blocks}. Table \tableautorefname~\ref{table:ablation} investigates the effect of activation layers in SNeRV blocks. By replacing the SINE layer with a common GELU layer (Variant V4), we find that the periodic inductive bias of the SINE layer is beneficial to the reconstruction quality. Subsequently, we explore different convolution configurations. As shown in \tableautorefname~\ref{table:block_ablation} and Fig.~\ref{fig:params}, integrating SINE in the NeRV, E-NeRV, and FFNeRV blocks results in a significant performance improvement of about 0.7dB. However, only modest improvements are observed in the HNeRV and SNeRV modules. This is attributed to the smaller STD values and more evenly distributed parameters within the HNeRV and SNeRV modules, helping the layers near the output to have sufficient capacity to store high-resolution video content and details. Thus, these modules efficiently fit videos without requiring elaborate feature extraction.

\noindent \textbf{Loss Function}. Different loss functions are compared in \tableautorefname~\ref{table:ablation}. These results demonstrate that integrating the MS-SSIM and frequency domain losses into L1 loss effectively enhances the quality of video reconstruction.

% \noindent \textbf{Loss Function}. We compare different loss functions in \tableautorefname~\ref{table:ablation}. By incorporating the MS-SSIM and frequency domain loss into L1 loss, our augmented INRs can effectively store the high-frequency information.

\begin{figure}[t]
    \centering
    \includegraphics[scale=0.3]{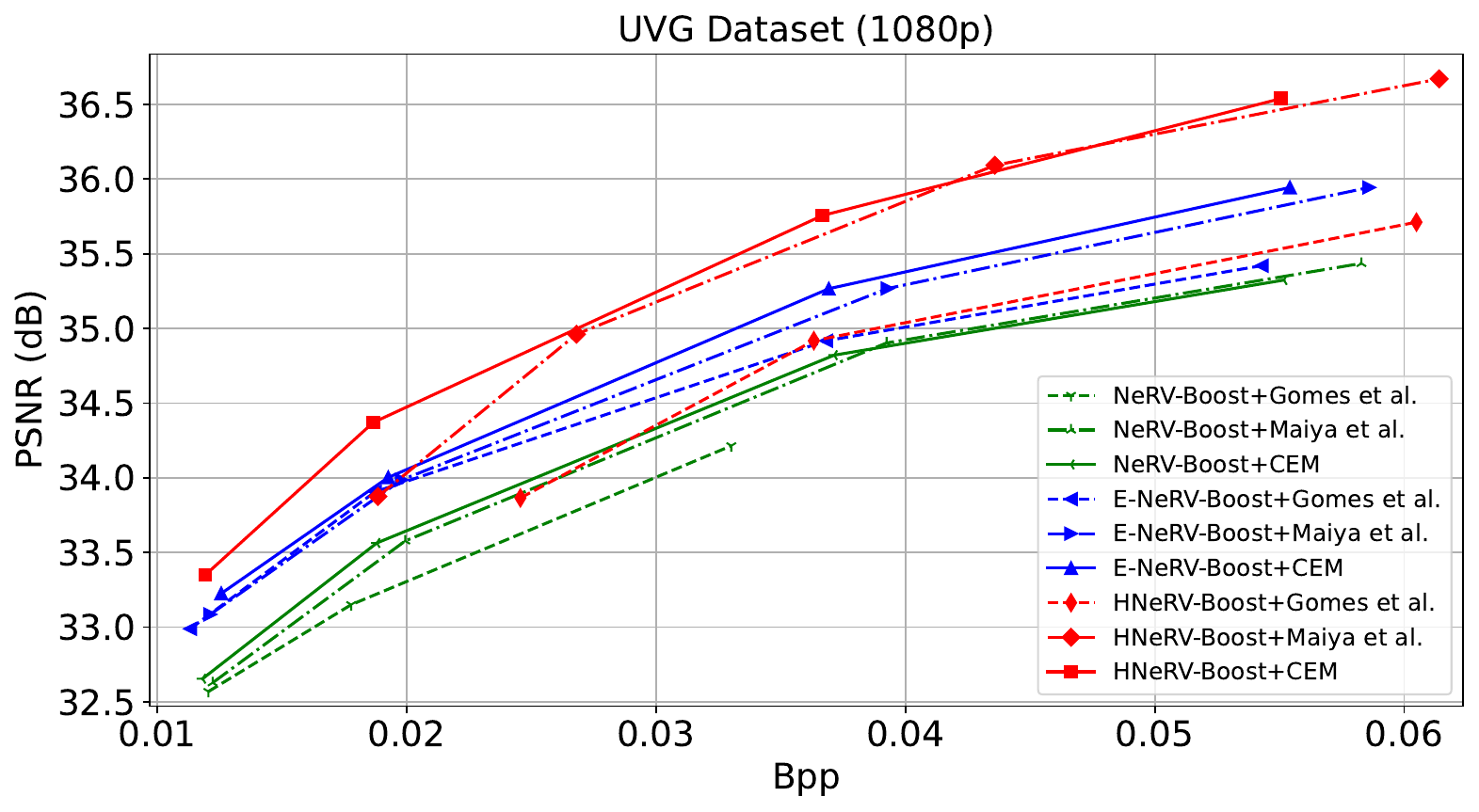}
     \vspace{-0.3cm}
    \caption{Rate-distortion comparisons between different entropy minimization techniques on the UVG dataset in PSNR.}
    \label{fig:em}
    \vspace{-0.5cm}
\end{figure}

% (Consistent Entropy Minimization) 
\noindent \textbf{Entropy Minimization}. To highlight the contribution of our proposed CEM scheme, Fig.~\ref{fig:em} displays the RD curves comparing various entropy minimization techniques on our boosted INR models. Notably, for content-dependent embedding compression in the boosted HNeRV, all methods employ asymmetric quantization. Since Gomes \textit{et al.} \cite{gomes2023video} use the asymmetric quantization for model weights, it tends to shift the quantized weights away from their original distribution, causing inferior RD results.  Compared with Maiya \textit{et al.} \cite{maiya2023nirvana}, our CEM method achieves more bitrate savings by maintaining consistency in the entropy model between training and inference. These results verify the superiority of the CEM technique in INR compression.

\section{Conclusion}
\label{sec:conclusion}
In this paper, we develop a universal framework to boost implicit video representations, achieving substantial improvements in key tasks like regression, compression, inpainting, and interpolation. These advancements are primarily due to the integration of several novel developments, including the temporal-aware affine transform, sinusoidal NeRV-like block design, improved reconstruction loss, and consistent entropy minimization. Through comprehensive evaluations against multiple implicit video models, our boosted models demonstrate superior performance, setting a new benchmark in the field of implicit video representation. The contribution of each component is validated through extensive ablation studies. 
%In the future, we will apply our boosting framework to more implicit models to tap their potential.

\section*{Acknowledgments}
This work was supported by the General Research Fund (Project No. 16209622) from the Hong Kong Research Grants Council.

{
    \small
    \bibliographystyle{ieeenat_fullname}
    \bibliography{main}

\begin{thebibliography}{55}
\providecommand{\natexlab}[1]{#1}
\providecommand{\url}[1]{\texttt{#1}}
\expandafter\ifx\csname urlstyle\endcsname\relax
  \providecommand{\doi}[1]{doi: #1}\else
  \providecommand{\doi}{doi: \begingroup \urlstyle{rm}\Url}\fi

\bibitem[Bai et~al.(2023)Bai, Dong, Wang, and Yuan]{bai2023ps}
Yunpeng Bai, Chao Dong, Cairong Wang, and Chun Yuan.
\newblock Ps-nerv: Patch-wise stylized neural representations for videos.
\newblock In \emph{2023 IEEE International Conference on Image Processing (ICIP)}, pages 41--45. IEEE, 2023.

\bibitem[Ball{\'e} et~al.(2017)Ball{\'e}, Laparra, and Simoncelli]{balle2017end}
Johannes Ball{\'e}, Valero Laparra, and Eero~P Simoncelli.
\newblock End-to-end optimized image compression.
\newblock In \emph{5th International Conference on Learning Representations, ICLR 2017}, 2017.

\bibitem[Ball{\'e} et~al.(2018)Ball{\'e}, Minnen, Singh, Hwang, and Johnston]{balle2018variational}
Johannes Ball{\'e}, David Minnen, Saurabh Singh, Sung~Jin Hwang, and Nick Johnston.
\newblock Variational image compression with a scale hyperprior.
\newblock In \emph{International Conference on Learning Representations}, 2018.

\bibitem[Bamler(2022)]{bamler2022constriction}
Robert Bamler.
\newblock Understanding entropy coding with asymmetric numeral systems (ans): a statistician's perspective.
\newblock \emph{arXiv preprint arXiv:2201.01741}, 2022.

\bibitem[B{\'e}gaint et~al.(2020)B{\'e}gaint, Racap{\'e}, Feltman, and Pushparaja]{begaint2020compressai}
Jean B{\'e}gaint, Fabien Racap{\'e}, Simon Feltman, and Akshay Pushparaja.
\newblock Compressai: a pytorch library and evaluation platform for end-to-end compression research.
\newblock \emph{arXiv preprint arXiv:2011.03029}, 2020.

\bibitem[Bhalgat et~al.(2020)Bhalgat, Lee, Nagel, Blankevoort, and Kwak]{bhalgat2020lsq+}
Yash Bhalgat, Jinwon Lee, Markus Nagel, Tijmen Blankevoort, and Nojun Kwak.
\newblock Lsq+: Improving low-bit quantization through learnable offsets and better initialization.
\newblock In \emph{Proceedings of the IEEE/CVF Conference on Computer Vision and Pattern Recognition Workshops}, pages 696--697, 2020.

\bibitem[Chen et~al.(2021)Chen, He, Wang, Ren, Lim, and Shrivastava]{chen2021nerv}
Hao Chen, Bo He, Hanyu Wang, Yixuan Ren, Ser~Nam Lim, and Abhinav Shrivastava.
\newblock Nerv: Neural representations for videos.
\newblock \emph{Advances in Neural Information Processing Systems}, 34:\penalty0 21557--21568, 2021.

\bibitem[Chen et~al.(2022)Chen, Gwilliam, He, Lim, and Shrivastava]{chen2022cnerv}
Hao Chen, Matt Gwilliam, Bo He, Ser-Nam Lim, and Abhinav Shrivastava.
\newblock Cnerv: Content-adaptive neural representation for visual data.
\newblock In \emph{British Machine Vision Conference}, 2022.

\bibitem[Chen et~al.(2023)Chen, Gwilliam, Lim, and Shrivastava]{chen2023hnerv}
Hao Chen, Matthew Gwilliam, Ser-Nam Lim, and Abhinav Shrivastava.
\newblock Hnerv: A hybrid neural representation for videos.
\newblock In \emph{Proceedings of the IEEE/CVF Conference on Computer Vision and Pattern Recognition}, pages 10270--10279, 2023.

\bibitem[Cheng et~al.(2020)Cheng, Sun, Takeuchi, and Katto]{cheng2020learned}
Zhengxue Cheng, Heming Sun, Masaru Takeuchi, and Jiro Katto.
\newblock Learned image compression with discretized gaussian mixture likelihoods and attention modules.
\newblock In \emph{Proceedings of the IEEE/CVF conference on computer vision and pattern recognition}, pages 7939--7948, 2020.

\bibitem[Dumoulin et~al.(2016)Dumoulin, Shlens, and Kudlur]{dumoulin2016learned}
Vincent Dumoulin, Jonathon Shlens, and Manjunath Kudlur.
\newblock A learned representation for artistic style.
\newblock In \emph{International Conference on Learning Representations}, 2016.

\bibitem[Esser et~al.(2019)Esser, McKinstry, Bablani, Appuswamy, and Modha]{esser2019learned}
Steven~K Esser, Jeffrey~L McKinstry, Deepika Bablani, Rathinakumar Appuswamy, and Dharmendra~S Modha.
\newblock Learned step size quantization.
\newblock In \emph{International Conference on Learning Representations}, 2019.

\bibitem[Ghiasi et~al.(2017)Ghiasi, Lee, Kudlur, Dumoulin, and Shlens]{ghiasi2017exploring}
Golnaz Ghiasi, Honglak Lee, Manjunath Kudlur, Vincent Dumoulin, and Jonathon Shlens.
\newblock Exploring the structure of a real-time, arbitrary neural artistic stylization network.
\newblock In \emph{British Machine Vision Conference}, 2017.

\bibitem[Gomes et~al.(2023)Gomes, Azevedo, and Schroers]{gomes2023video}
Carlos Gomes, Roberto Azevedo, and Christopher Schroers.
\newblock Video compression with entropy-constrained neural representations.
\newblock In \emph{Proceedings of the IEEE/CVF Conference on Computer Vision and Pattern Recognition}, pages 18497--18506, 2023.

\bibitem[He et~al.(2023)He, Yang, Wang, Wu, Chen, Huang, Ren, Lim, and Shrivastava]{he2023towards}
Bo He, Xitong Yang, Hanyu Wang, Zuxuan Wu, Hao Chen, Shuaiyi Huang, Yixuan Ren, Ser-Nam Lim, and Abhinav Shrivastava.
\newblock Towards scalable neural representation for diverse videos.
\newblock In \emph{Proceedings of the IEEE/CVF Conference on Computer Vision and Pattern Recognition}, pages 6132--6142, 2023.

\bibitem[Hu et~al.(2019)Hu, Li, Huang, and Gao]{hu2019channel}
Yanting Hu, Jie Li, Yuanfei Huang, and Xinbo Gao.
\newblock Channel-wise and spatial feature modulation network for single image super-resolution.
\newblock \emph{IEEE Transactions on Circuits and Systems for Video Technology}, 30\penalty0 (11):\penalty0 3911--3927, 2019.

\bibitem[Huang and Belongie(2017)]{huang2017arbitrary}
Xun Huang and Serge Belongie.
\newblock Arbitrary style transfer in real-time with adaptive instance normalization.
\newblock In \emph{Proceedings of the IEEE international conference on computer vision}, pages 1501--1510, 2017.

\bibitem[Kim et~al.(2020)Kim, Soh, Park, and Cho]{kim2020transfer}
Yoonsik Kim, Jae~Woong Soh, Gu~Yong Park, and Nam~Ik Cho.
\newblock Transfer learning from synthetic to real-noise denoising with adaptive instance normalization.
\newblock In \emph{Proceedings of the IEEE/CVF conference on computer vision and pattern recognition}, pages 3482--3492, 2020.

\bibitem[Kristensen(2010)]{bunny2010}
Janus~B. Kristensen.
\newblock Big buck bunny.
\newblock 2010.

\bibitem[Kwan et~al.(2023)Kwan, Gao, Zhang, Gower, and Bull]{kwan2023hinerv}
Ho~Man Kwan, Ge Gao, Fan Zhang, Andrew Gower, and David Bull.
\newblock Hinerv: Video compression with hierarchical encoding based neural representation.
\newblock \emph{arXiv preprint arXiv:2306.09818}, 2023.

\bibitem[Lee et~al.(2023)Lee, Rho, Ko, and Park]{lee2023ffnerv}
Joo~Chan Lee, Daniel Rho, Jong~Hwan Ko, and Eunbyung Park.
\newblock Ffnerv: Flow-guided frame-wise neural representations for videos.
\newblock In \emph{Proceedings of the ACM International Conference on Multimedia}, 2023.

\bibitem[Li et~al.(2021)Li, Li, and Lu]{li2021deep}
Jiahao Li, Bin Li, and Yan Lu.
\newblock Deep contextual video compression.
\newblock \emph{Advances in Neural Information Processing Systems}, 34:\penalty0 18114--18125, 2021.

\bibitem[Li et~al.(2022)Li, Wang, Pi, Xu, Mei, and Liu]{li2022nerv}
Zizhang Li, Mengmeng Wang, Huaijin Pi, Kechun Xu, Jianbiao Mei, and Yong Liu.
\newblock E-nerv: Expedite neural video representation with disentangled spatial-temporal context.
\newblock In \emph{European Conference on Computer Vision}, pages 267--284. Springer, 2022.

\bibitem[Li et~al.(2023)Li, Wang, and Meng]{li2023regularize}
Zhemin Li, Hongxia Wang, and Deyu Meng.
\newblock Regularize implicit neural representation by itself.
\newblock In \emph{Proceedings of the IEEE/CVF Conference on Computer Vision and Pattern Recognition}, pages 10280--10288, 2023.

\bibitem[Lin et~al.(2020)Lin, Liu, Li, and Wu]{lin2020m}
Jianping Lin, Dong Liu, Houqiang Li, and Feng Wu.
\newblock M-lvc: Multiple frames prediction for learned video compression.
\newblock In \emph{Proceedings of the IEEE/CVF Conference on Computer Vision and Pattern Recognition}, pages 3546--3554, 2020.

\bibitem[Liu et~al.(2022)Liu, Mao, Wu, Feichtenhofer, Darrell, and Xie]{liu2022convnet}
Zhuang Liu, Hanzi Mao, Chao-Yuan Wu, Christoph Feichtenhofer, Trevor Darrell, and Saining Xie.
\newblock A convnet for the 2020s.
\newblock In \emph{Proceedings of the IEEE/CVF conference on computer vision and pattern recognition}, pages 11976--11986, 2022.

\bibitem[Lu et~al.(2019)Lu, Ouyang, Xu, Zhang, Cai, and Gao]{lu2019dvc}
Guo Lu, Wanli Ouyang, Dong Xu, Xiaoyun Zhang, Chunlei Cai, and Zhiyong Gao.
\newblock Dvc: An end-to-end deep video compression framework.
\newblock In \emph{Proceedings of the IEEE/CVF Conference on Computer Vision and Pattern Recognition}, pages 11006--11015, 2019.

\bibitem[Maiya et~al.(2023)Maiya, Girish, Ehrlich, Wang, Lee, Poirson, Wu, Wang, and Shrivastava]{maiya2023nirvana}
Shishira~R Maiya, Sharath Girish, Max Ehrlich, Hanyu Wang, Kwot~Sin Lee, Patrick Poirson, Pengxiang Wu, Chen Wang, and Abhinav Shrivastava.
\newblock Nirvana: Neural implicit representations of videos with adaptive networks and autoregressive patch-wise modeling.
\newblock In \emph{Proceedings of the IEEE/CVF Conference on Computer Vision and Pattern Recognition}, pages 14378--14387, 2023.

\bibitem[Mercat et~al.(2020)Mercat, Viitanen, and Vanne]{mercat2020uvg}
Alexandre Mercat, Marko Viitanen, and Jarno Vanne.
\newblock Uvg dataset: 50/120fps 4k sequences for video codec analysis and development.
\newblock In \emph{Proceedings of the 11th ACM Multimedia Systems Conference}, pages 297--302, 2020.

\bibitem[Mildenhall et~al.(2021)Mildenhall, Srinivasan, Tancik, Barron, Ramamoorthi, and Ng]{mildenhall2021nerf}
Ben Mildenhall, Pratul~P Srinivasan, Matthew Tancik, Jonathan~T Barron, Ravi Ramamoorthi, and Ren Ng.
\newblock Nerf: Representing scenes as neural radiance fields for view synthesis.
\newblock \emph{Communications of the ACM}, 65\penalty0 (1):\penalty0 99--106, 2021.

\bibitem[M{\"u}ller et~al.(2022)M{\"u}ller, Evans, Schied, and Keller]{muller2022instant}
Thomas M{\"u}ller, Alex Evans, Christoph Schied, and Alexander Keller.
\newblock Instant neural graphics primitives with a multiresolution hash encoding.
\newblock \emph{ACM Transactions on Graphics (ToG)}, 41\penalty0 (4):\penalty0 1--15, 2022.

\bibitem[Oktay et~al.(2019)Oktay, Ball{\'e}, Singh, and Shrivastava]{oktay2019scalable}
Deniz Oktay, Johannes Ball{\'e}, Saurabh Singh, and Abhinav Shrivastava.
\newblock Scalable model compression by entropy penalized reparameterization.
\newblock In \emph{International Conference on Learning Representations}, 2019.

\bibitem[Park et~al.(2021)Park, Sinha, Barron, Bouaziz, Goldman, Seitz, and Martin-Brualla]{park2021nerfies}
Keunhong Park, Utkarsh Sinha, Jonathan~T Barron, Sofien Bouaziz, Dan~B Goldman, Steven~M Seitz, and Ricardo Martin-Brualla.
\newblock Nerfies: Deformable neural radiance fields.
\newblock In \emph{Proceedings of the IEEE/CVF International Conference on Computer Vision}, pages 5865--5874, 2021.

\bibitem[Park et~al.(2019)Park, Liu, Wang, and Zhu]{park2019semantic}
Taesung Park, Ming-Yu Liu, Ting-Chun Wang, and Jun-Yan Zhu.
\newblock Semantic image synthesis with spatially-adaptive normalization.
\newblock In \emph{Proceedings of the IEEE/CVF conference on computer vision and pattern recognition}, pages 2337--2346, 2019.

\bibitem[Perazzi et~al.(2016)Perazzi, Pont-Tuset, McWilliams, Van~Gool, Gross, and Sorkine-Hornung]{perazzi2016benchmark}
Federico Perazzi, Jordi Pont-Tuset, Brian McWilliams, Luc Van~Gool, Markus Gross, and Alexander Sorkine-Hornung.
\newblock A benchmark dataset and evaluation methodology for video object segmentation.
\newblock In \emph{Proceedings of the IEEE conference on computer vision and pattern recognition}, pages 724--732, 2016.

\bibitem[Perez et~al.(2018)Perez, Strub, De~Vries, Dumoulin, and Courville]{perez2018film}
Ethan Perez, Florian Strub, Harm De~Vries, Vincent Dumoulin, and Aaron Courville.
\newblock Film: Visual reasoning with a general conditioning layer.
\newblock In \emph{Proceedings of the AAAI conference on artificial intelligence}, 2018.

\bibitem[Pumarola et~al.(2021)Pumarola, Corona, Pons-Moll, and Moreno-Noguer]{pumarola2021d}
Albert Pumarola, Enric Corona, Gerard Pons-Moll, and Francesc Moreno-Noguer.
\newblock D-nerf: Neural radiance fields for dynamic scenes.
\newblock In \emph{Proceedings of the IEEE/CVF Conference on Computer Vision and Pattern Recognition}, pages 10318--10327, 2021.

\bibitem[Saragadam et~al.(2023)Saragadam, LeJeune, Tan, Balakrishnan, Veeraraghavan, and Baraniuk]{saragadam2023wire}
Vishwanath Saragadam, Daniel LeJeune, Jasper Tan, Guha Balakrishnan, Ashok Veeraraghavan, and Richard~G Baraniuk.
\newblock Wire: Wavelet implicit neural representations.
\newblock In \emph{Proceedings of the IEEE/CVF Conference on Computer Vision and Pattern Recognition}, pages 18507--18516, 2023.

\bibitem[Sheng et~al.(2022)Sheng, Li, Li, Li, Liu, and Lu]{sheng2022temporal}
Xihua Sheng, Jiahao Li, Bin Li, Li Li, Dong Liu, and Yan Lu.
\newblock Temporal context mining for learned video compression.
\newblock \emph{IEEE Transactions on Multimedia}, 2022.

\bibitem[Sitzmann et~al.(2020)Sitzmann, Martel, Bergman, Lindell, and Wetzstein]{sitzmann2020implicit}
Vincent Sitzmann, Julien Martel, Alexander Bergman, David Lindell, and Gordon Wetzstein.
\newblock Implicit neural representations with periodic activation functions.
\newblock \emph{Advances in neural information processing systems}, 33:\penalty0 7462--7473, 2020.

\bibitem[Skorokhodov et~al.(2021)Skorokhodov, Ignatyev, and Elhoseiny]{skorokhodov2021adversarial}
Ivan Skorokhodov, Savva Ignatyev, and Mohamed Elhoseiny.
\newblock Adversarial generation of continuous images.
\newblock In \emph{Proceedings of the IEEE/CVF conference on computer vision and pattern recognition}, pages 10753--10764, 2021.

\bibitem[Skorokhodov et~al.(2022)Skorokhodov, Tulyakov, and Elhoseiny]{skorokhodov2022stylegan}
Ivan Skorokhodov, Sergey Tulyakov, and Mohamed Elhoseiny.
\newblock Stylegan-v: A continuous video generator with the price, image quality and perks of stylegan2.
\newblock In \emph{Proceedings of the IEEE/CVF Conference on Computer Vision and Pattern Recognition}, pages 3626--3636, 2022.

\bibitem[Song et~al.(2021)Song, Choi, and Han]{song2021variable}
Myungseo Song, Jinyoung Choi, and Bohyung Han.
\newblock Variable-rate deep image compression through spatially-adaptive feature transform.
\newblock In \emph{Proceedings of the IEEE/CVF International Conference on Computer Vision}, pages 2380--2389, 2021.

\bibitem[Str{\"u}mpler et~al.(2022)Str{\"u}mpler, Postels, Yang, Gool, and Tombari]{strumpler2022implicit}
Yannick Str{\"u}mpler, Janis Postels, Ren Yang, Luc~Van Gool, and Federico Tombari.
\newblock Implicit neural representations for image compression.
\newblock In \emph{European Conference on Computer Vision}, pages 74--91. Springer, 2022.

\bibitem[Sullivan et~al.(2012)Sullivan, Ohm, Han, and Wiegand]{sullivan2012overview}
Gary~J Sullivan, Jens-Rainer Ohm, Woo-Jin Han, and Thomas Wiegand.
\newblock Overview of the high efficiency video coding (hevc) standard.
\newblock \emph{IEEE Transactions on circuits and systems for video technology}, 22\penalty0 (12):\penalty0 1649--1668, 2012.

\bibitem[Szatkowski et~al.(2022)Szatkowski, Piczak, Spurek, Tabor, and Trzci{\'n}ski]{szatkowski2022hypersound}
Filip Szatkowski, Karol~J Piczak, Przemys{\l}aw Spurek, Jacek Tabor, and Tomasz Trzci{\'n}ski.
\newblock Hypersound: Generating implicit neural representations of audio signals with hypernetworks.
\newblock \emph{arXiv preprint arXiv:2211.01839}, 2022.

\bibitem[Tang et~al.(2023)Tang, Zhang, Zhang, and Ma]{tang2023scene}
Lv Tang, Xinfeng Zhang, Gai Zhang, and Xiaoqi Ma.
\newblock Scene matters: Model-based deep video compression.
\newblock In \emph{Proceedings of the IEEE international conference on computer vision}, 2023.

\bibitem[Wang et~al.(2018)Wang, Yu, Dong, and Loy]{wang2018recovering}
Xintao Wang, Ke Yu, Chao Dong, and Chen~Change Loy.
\newblock Recovering realistic texture in image super-resolution by deep spatial feature transform.
\newblock In \emph{Proceedings of the IEEE conference on computer vision and pattern recognition}, pages 606--615, 2018.

\bibitem[Wang et~al.(2021)Wang, Li, Zhang, and Shan]{wang2021towards}
Xintao Wang, Yu Li, Honglun Zhang, and Ying Shan.
\newblock Towards real-world blind face restoration with generative facial prior.
\newblock In \emph{Proceedings of the IEEE/CVF conference on computer vision and pattern recognition}, pages 9168--9178, 2021.

\bibitem[Wiegand et~al.(2003)Wiegand, Sullivan, Bjontegaard, and Luthra]{wiegand2003overview}
Thomas Wiegand, Gary~J Sullivan, Gisle Bjontegaard, and Ajay Luthra.
\newblock Overview of the h. 264/avc video coding standard.
\newblock \emph{IEEE Transactions on circuits and systems for video technology}, 13\penalty0 (7):\penalty0 560--576, 2003.

\bibitem[Xie et~al.(2022)Xie, Zhou, Li, Lin, and Shuicheng]{xie2022adan}
Xingyu Xie, Pan Zhou, Huan Li, Zhouchen Lin, and YAN Shuicheng.
\newblock Adan: Adaptive nesterov momentum algorithm for faster optimizing deep models.
\newblock In \emph{Has it Trained Yet? NeurIPS 2022 Workshop}, 2022.

\bibitem[Xu and Jiao(2023)]{xu2023revisiting}
Wentian Xu and Jianbo Jiao.
\newblock Revisiting implicit neural representations in low-level vision.
\newblock 2023.

\bibitem[Yang et~al.(2023)Yang, Xiao, Cheng, Cao, Qu, Suo, and Dai]{yang2023sci}
Runzhao Yang, Tingxiong Xiao, Yuxiao Cheng, Qianni Cao, Jinyuan Qu, Jinli Suo, and Qionghai Dai.
\newblock Sci: A spectrum concentrated implicit neural compression for biomedical data.
\newblock In \emph{Proceedings of the AAAI Conference on Artificial Intelligence}, pages 4774--4782, 2023.

\bibitem[Zhang et~al.(2022)Zhang, van Rozendaal, Brehmer, Nagel, and Cohen]{zhang2022implicit}
Yunfan Zhang, Ties van Rozendaal, Johann Brehmer, Markus Nagel, and Taco Cohen.
\newblock Implicit neural video compression.
\newblock In \emph{ICLR Workshop on Deep Generative Models for Highly Structured Data}, 2022.

\bibitem[Zhao et~al.(2023)Zhao, Asif, and Ma]{zhao2023dnerv}
Qi Zhao, M~Salman Asif, and Zhan Ma.
\newblock Dnerv: Modeling inherent dynamics via difference neural representation for videos.
\newblock In \emph{Proceedings of the IEEE/CVF Conference on Computer Vision and Pattern Recognition}, pages 2031--2040, 2023.

\end{thebibliography}
}

\clearpage
\appendix
% \setcounter{page}{1}
% \maketitlesupplementary

\section{Experimental Results}
\begin{figure*}[t]
    \centering
    \includegraphics[scale=0.84]{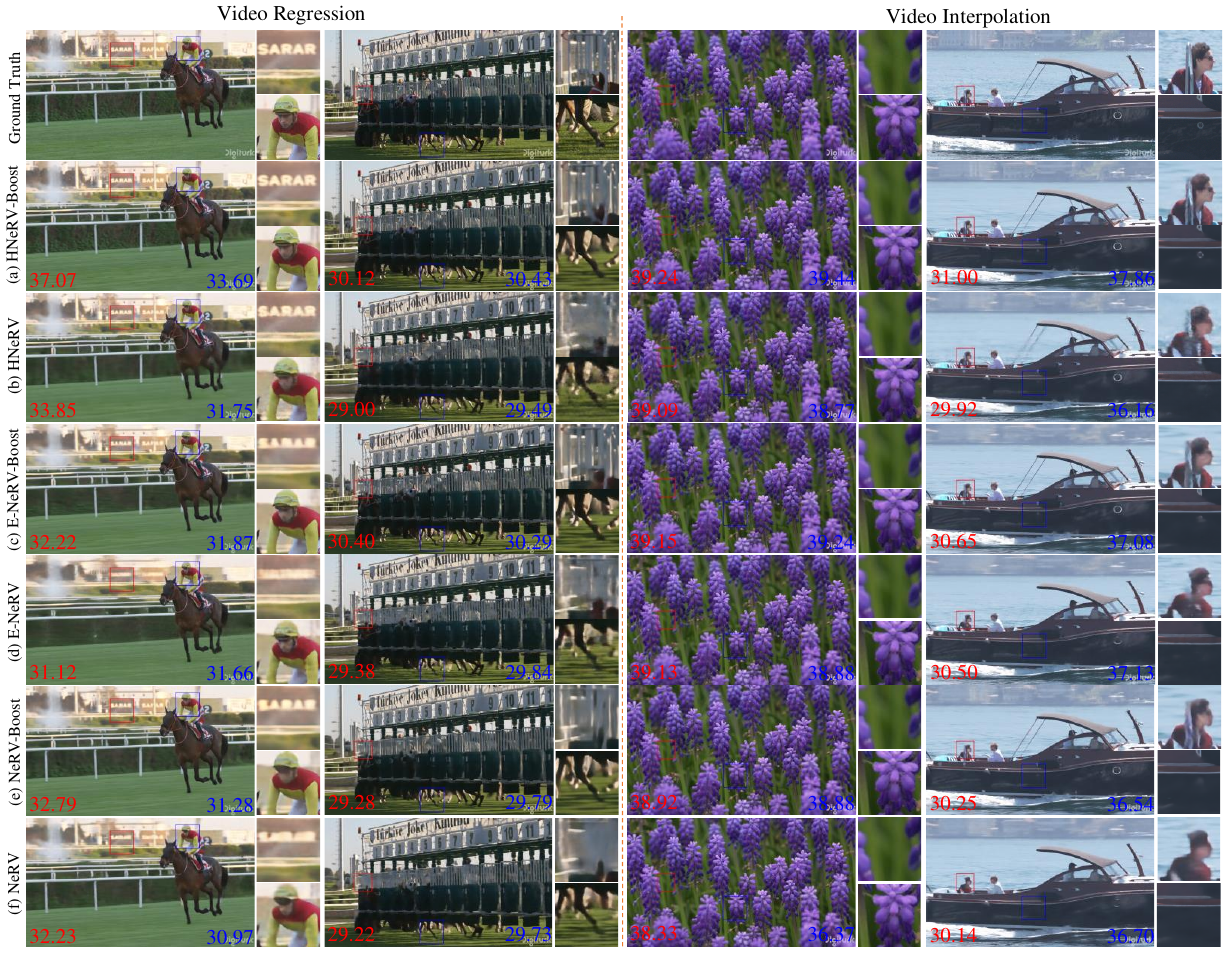}
    \caption{Visualization comparison of regression and interpolation on the UVG dataset. The top row displays the ground truth, followed by our boosted results in (a, c, e) and the baseline results in (b, d, f). The red and blue numbers indicate the PSNR values for the respective colored patches. }
    \label{fig:visualizations_reg_inter}
    \vspace{-0.3cm}
\end{figure*}

\begin{figure*}[t]
    \centering
    \includegraphics[scale=0.84]{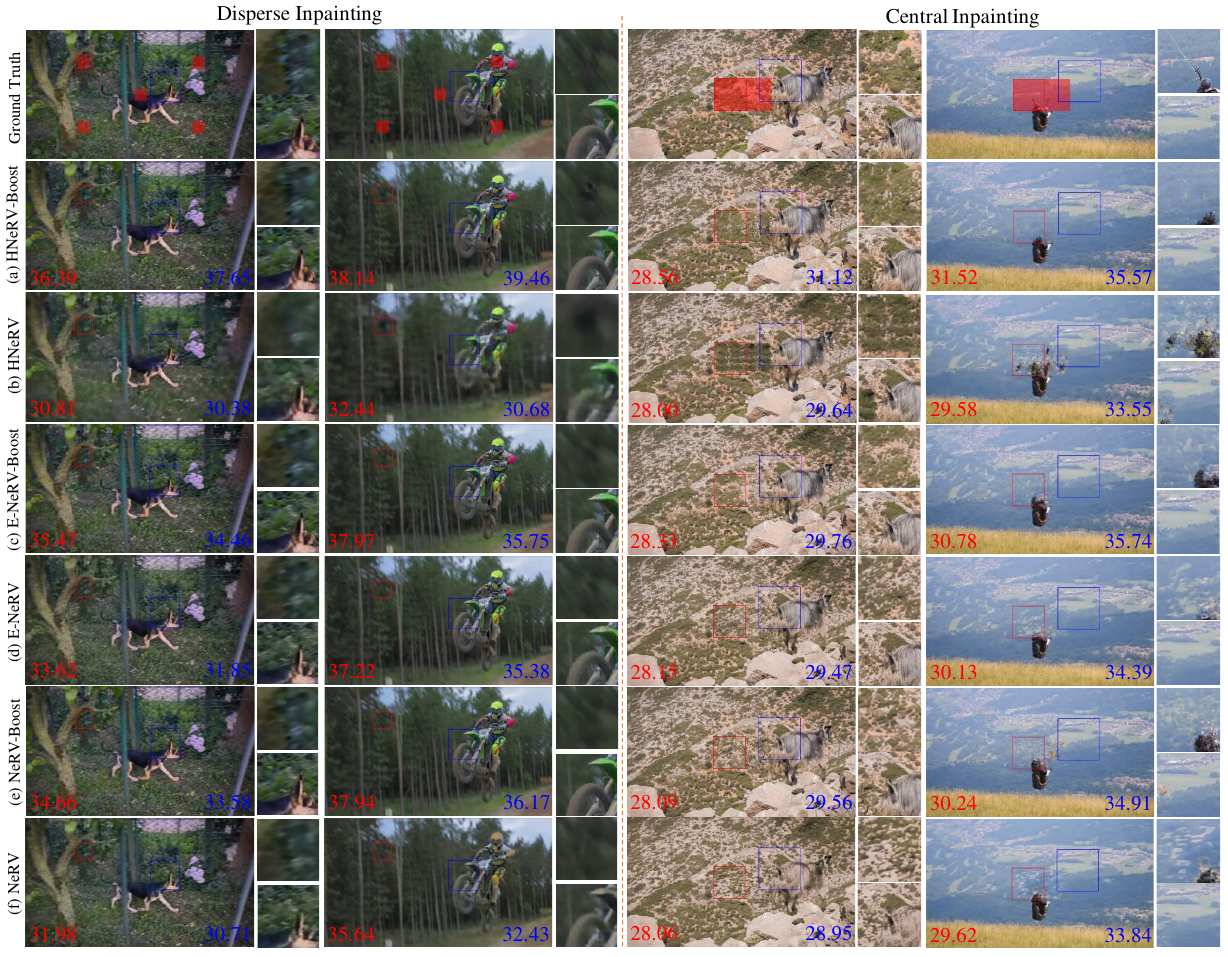}
    \caption{Visualization comparison of disperse and central inpainting on the DAVIS validation dataset. The top row displays the ground truth, followed by our boosted results in (a, c, e) and the baseline results in (b, d, f). The red and blue numbers indicate the PSNR values for the respective colored patches.}
    \label{fig:visualizations_inpaint}
    \vspace{-0.3cm}
\end{figure*}

\noindent \textbf{Video Regression, Interpolation and Inpainting}. For video regression, we provide the complete PSNR and MS-SSIM results on the Buuny and UVG datasets in \tableautorefname~\ref{table:size_bunny_all} and \tableautorefname~\ref{table:uvg_details_all}. Besides, MS-SSIM results for video interpolation and inpainting on the UVG and DAVIS validation datasets are reported in \tableautorefname~\ref{table:uvg_interpolation_msssim} and \tableautorefname~\ref{table:davis_details_msssim}, respectively. 
Further visualizations are available in Fig.~\ref{fig:visualizations_reg_inter} and Fig.~\ref{fig:visualizations_inpaint}.

\noindent \textbf{Encoding Time}. The encoding time comparison for video regression is shown in \tableautorefname~\ref{table:encoding_complexity}. Our enhanced models exhibit about a twofold increase in encoding latency.

% We provide the encoding time comparison for video regression in \tableautorefname~\ref{table:encoding_complexity}. Our boosted models only paying 2x encoding latency.
\noindent \textbf{Additional baselines.} Apart from the classic NeRV, E-NeRV and HNeRV, our boosted method can also apply to recently proposed implicit video models. \tableautorefname~\ref{table:baselines} reports the regression results of DNeRV \cite{he2023towards}, DivNeRV \cite{zhao2023dnerv} and their boosted versions on the UVG dataset. For DNeRV, we follow its setting to crop videos into 1024$\times$1920. The results show that our approach boosts DNeRV and DivNeRV by 0.77dB and 1.42dB PSNR, respectively.

\noindent \textbf{Ablation Study}. To further verify the contribution of each boosting component, a set of ablation experiments are conducted in \tableautorefname~\ref{table:ablation_components}.
Since NeRV, E-NeRV, and HNeRV have an equal number of parameters, they shares similar upper bounds in model fitting capabilities. 
Compared with original INR baselines, the integration of temporal-aware affine transform (TAT) allows intermediate features to effectively align with the target frame, enhancing the reconstruction quality.
Subsequently, we introduce a sinusoidal NeRV-like (SNeRV) block, facilitating diverse feature generation and balanced parameter distribution.
Unlike NeRV and E-NeRV, which use an index for frame generation, HNeRV introduces content-aware embedding as prior information for each frame, facilitating efficient video fitting without elaborate feature extraction. Without the help of prior information, the SNeRV block significantly enhances NeRV and E-NeRV's performance by 2.78dB and 1.33dB, respectively.
We further incorporate a high-frequency information-preserving loss that enables INRs to capture intricate video details more effectively.

% Compared with original INR baselines, the integration of temporal-aware affine transform (TAT) allows intermediate features to effectively align with the target frame, enhancing the reconstruction quality. Subsequently, we introduce a sinusoidal NeRV-like (SNeRV) block, facilitating diverse feature generation and balanced parameter distribution, which has proven effective. We further incorporate a high-frequency information-preserving loss that enables INRs to capture intricate video details more effectively.

% \tableautorefname~5 of the appendix shows that NeRV, E-NeRV, and HNeRV have an equal number of parameters, sharing similar upper bounds in model fitting capabilities. Unlike NeRV and E-NeRV, which use an index for frame generation, HNeRV introduces content-aware embedding as prior information for each frame, facilitating efficient video fitting without elaborate feature extraction. Without the help of prior information, the SNeRV block significantly enhances NeRV and E-NeRV's performance by 2.78dB and 1.33dB, respectively.

\section{Video Compression Details}
To build video codecs, we fine-tune INR models using our consistent entropy minimization (CEM) method. For the entropy regularization term, we set $\kappa$ as 0.05, 0.2, and 0.5 in the NeRV/E-NeRV/HNeRV, NeRV-Boost/E-NeRV-Bosst, and HNeRV-Boost, respectively. We use asymmetric numeral systems of the constriction library \cite{bamler2022constriction} to perform entropy coding. In recent learning-based video codecs, we obtain the results of DCVC \cite{li2021deep} and DCVC-TCM \cite{sheng2022temporal} by employing their open source codes\footnote{\url{https://github.com/microsoft/DCVC/tree/main}} with a Group of Pictures (GOP) size of 32. The I frame codecs in DCVC utilize the pretrained models \textit{cheng-2020} \cite{cheng2020learned} with quality indices 3,4,5,6 as provided by CompressAI \cite{begaint2020compressai}. For traditional codecs, we implement H.264 and H.265 with the veryslow preset and enabling B frames by using the following commands:
\begin{itemize}
\item H.264: \textit{ffmpeg -pix\_fmt yuv420p -s:v W×H -i Video.yuv -vframes $N_e$ -c:v libx264 -preset veryslow -qp $QP$ -g $GOP$ output.mkv}
\item H.265: \textit{ffmpeg -pix\_fmt yuv420p -s:v W×H -i Video.yuv -vframes $N_e$ -c:v libx265 -preset veryslow -x265-params “qp=$QP$:keyint=$GOP$” output.mkv}
\end{itemize}
where $N_e$ denotes the number of encoded frames and $QP$ represents the number of quantization parameters. The GOP size is set as 32. 

\begin{figure}[t]
    \centering
    \includegraphics[scale=0.25]{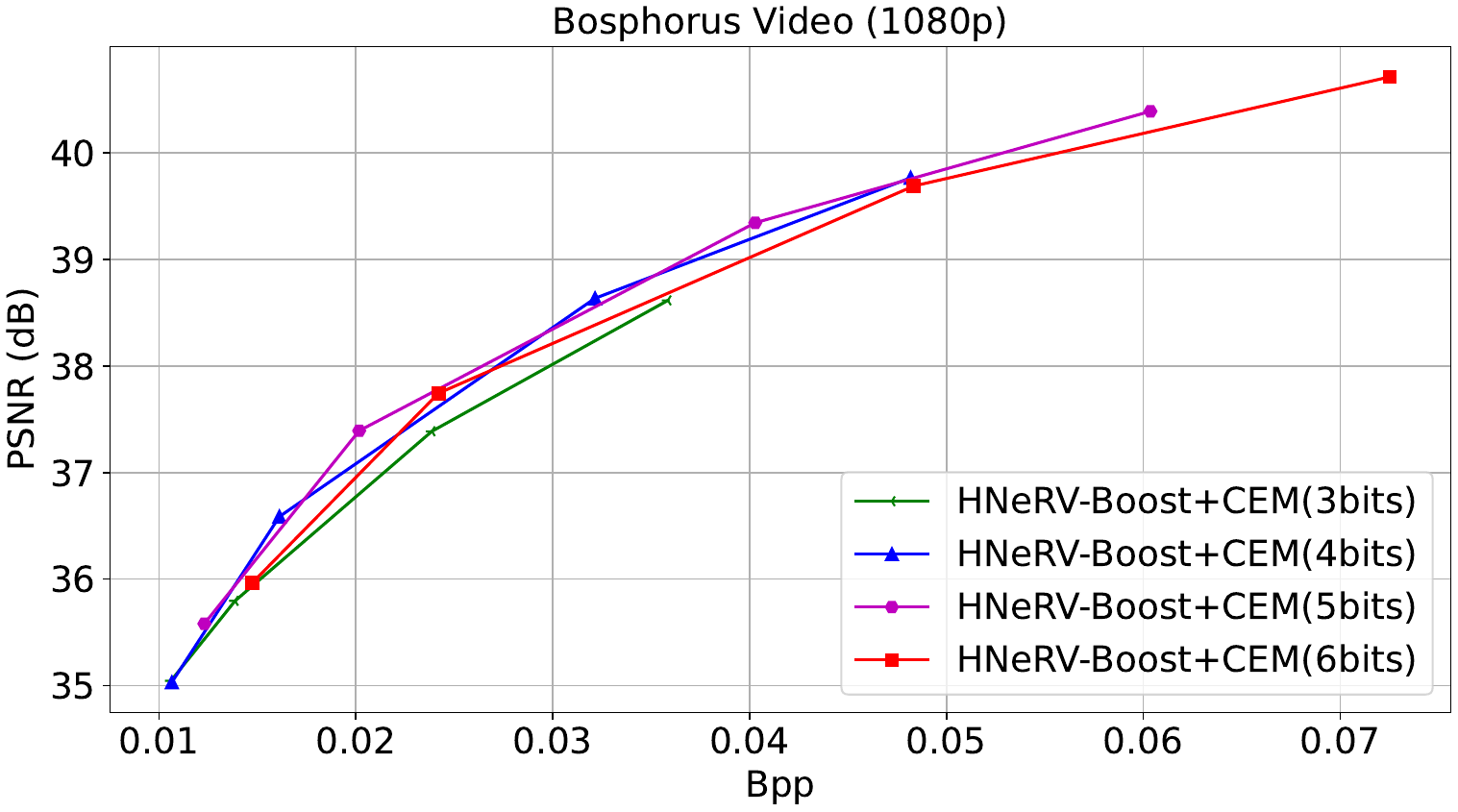}
     \vspace{-0.3cm}
    \caption{Comparisons between different $B_{avg}$ bits.}
    \label{fig:bosph_bits}
    \vspace{-0.3cm}
\end{figure}

\noindent \textbf{Bitrate computation.} For INR-based codecs, we model the probability $p(\boldsymbol{\hat{y}}_t)$ of the quantized embedding as a Gaussian distribution. Then the estimated bitrate $R(\boldsymbol{\hat{y}}_{t})$ of the quantized embedding can be expressed as:
\begin{equation}
\begin{aligned}
R(\boldsymbol{\hat{y}}_{t}) = \mathbb{E}_{\boldsymbol{x}\sim p_{\boldsymbol{x}}} [-\log_2 p(\boldsymbol{\hat{y}}_t)]
\end{aligned}
\end{equation}
The estimated bitrates $R(\boldsymbol{\hat{\theta}})$ and $R(\boldsymbol{\hat{\psi}})$ follow the above similar computation procedure. In addition, when optimizing the RD trade-off, we need to choose the suitable $B_{avg}$. 
Fig.~\ref{fig:bosph_bits} illustrates the rate-distortion curves for various $B_{avg}$ values on our boosted HNeRV. We see that $B_{avg}=$ 4 or 5 bits yield superior RD performance compared to other scenarios. Based on empirical evaluation, $B_{avg}$ is fixed as 4 bits.

\section{Network Structure}
\label{sec:network}
\begin{figure*}[t]
    \centering
    \includegraphics[scale=0.8]{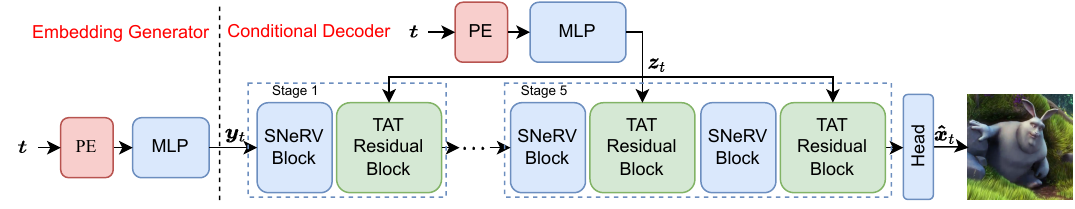}
    \caption{Our proposed NeRV-Boost framework with the conditional decoder. We follow the original NeRV to upsample the embedding $\boldsymbol{y}_t$ in stages 1 to 5.}
    \label{fig:nerv_arch}
    \vspace{-0.5cm}
\end{figure*}

\begin{figure*}[t]
    \centering
    \includegraphics[scale=0.72]{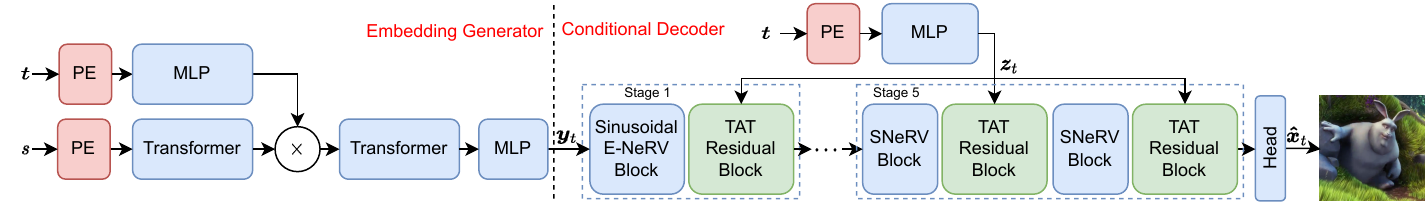}
    \caption{Our proposed E-NeRV-Boost framework with the conditional decoder. We follow the original E-NeRV to adopt the E-NeRV block in the stage 1, while the GELU activation is replaced with the SINE activation in the E-NeRV block. The embedding $\boldsymbol{y}_t$ is progressively upsampled and modulated in stages 1 to 5.}
    \label{fig:enerv_arch}
    \vspace{-0.4cm}
\end{figure*}

Fig.~\ref{fig:nerv_arch} and Fig.~\ref{fig:enerv_arch} illustrate our proposed NeRV-Boost and E-NeRV-Boost frameworks. Given NeRV and E-NeRV baselines, our approach involves three key steps. Firstly, inserting TAT residual blocks after NeRV-like blocks. Secondly, replacing the GELU layer with a SINE layer in NeRV-like blocks. Finally, in the last three stages, stacking two sinusoidal NeRV-like blocks for feature upsampling and refinement. Since NeRV and E-NeRV utilize 3$\times$3 kernels in their upsampling blocks, there is no need to alter the kernel size in the last three stages. These three revision steps effectively transform most video INR models into our INR-Boost models.

\tableautorefname~\ref{table:sub_modules} and \tableautorefname~\ref{table:conditional_decoder} give the network details of our condition decoders in different INR-Boost models. Our boosted models maintain the same channel reduction factor as their baseline counterparts. Model size variations are achieved by adjusting the channel width $C_1$. For the INR baselines, we employed their open-source codes, following specific training commands.
\begin{itemize}
\item NeRV\footnote{\url{https://github.com/haochen-rye/NeRV/tree/master}}: \textit{python train\_nerv.py -e 300 -{}-lower-width 12 -{}-num-blocks 1 -{}-dataset \{dataset\} -{}-frame\_gap 1 -{}-outf \{outf\} -{}-embed 1.25\_80 -{}-stem\_dim\_num 256\_1 -{}-reduction 2 -{}-fc\_hw\_dim 9\_16\_\{Depth\} -{}-expansion 1 -{}-single\_res -{}-loss Fusion6 -{}-warmup 0.2 -{}-lr\_type cosine -{}-strides 5 3 2 2 2 -{}-conv\_type conv -b 1 -{}-lr 0.0005 -{}-norm none -{}-act gelu}
\item E-NeRV\footnote{\url{https://github.com/kyleleey/E-NeRV/tree/main}}: \textit{python main.py -{}-output\_dir \{outf\} -{}-cfg\_path [stem\_dim\_num: 256, fc\_hw\_dim: 9\_16\_{Depth}, stride\_list: 5 3 2 2 2, lower\_width:12, block\_dim: 128, mlp\_dim: 64]} 
\item HNeRV\footnote{\url{https://github.com/haochen-rye/HNeRV}}: \textit{python train\_nerv\_all.py -{}-outf \{outf\}  -{}-data\_path \{dataset\} --vid \{dataset\_name\} -{}-conv\_type convnext pshuffel -{}-act gelu -{}-norm none --crop\_list 1080\_1920 -{}-resize\_list -1 -{}-loss L2  -{}-enc\_strds 5 3 2 2 2 -{}-enc\_dim 64\_16 -{}-dec\_strds 5 3 2 2 2 -{}-ks 0\_1\_5 -{}-reduce 1.2 -{}-modelsize \{Size\} -e 300 -{}-lower\_width 12 -b 1 -{}-lr 0.001}
\end{itemize}

% \section{Limitations}
% As illustrated in \tableautorefname~\ref{table:encoding_complexity}, a primary limitation of our boosting framework is its slow encoding speed. Typically, it requires several hours to overfit a video sequence for optimal encoding quality, a challenge also present in existing video INR models. This issue could potentially be mitigated by employing meta-learning to expedite the training of video INRs. Furthermore, our boosting framework is primarily tailored for enhancing NeRV-like models, and may not be as effective for non-NeRV-like models, such as HiNeRV \cite{kwan2023hinerv}.

\begin{table*}
\footnotesize
  \caption{Video regression results on the Bunny video in PSNR and MS-SSIM. The total size of HNeRV and HNeRV-Boost include the embedding size and the network size.}
  \label{table:size_bunny_all}
  \vspace{-0.3cm}
  \centering
  \begin{tabular}{c|cccccc}
    \hline
    Model &  Size & PSNR & PSNR Avg. & MS-SSIM & MS-SSIM Avg. \\
    \hline
    NeRV \cite{chen2021nerv} &  0.76M/1.49M/3.03M & 26.12/28.54/31.84 & 28.83 &   0.8551/0.9183/0.9633 & 0.9122\\
    NeRV-Boost & 0.76M/1.51M/3.00M & \textbf{30.25}/\textbf{33.71}/\textbf{37.25} & \textbf{33.73} &\textbf{0.9341}/\textbf{0.9690}/\textbf{0.9860} & \textbf{0.9630}\\
    \hline
    E-NeRV \cite{li2022nerv}& 0.76M/1.51M/3.00M & 29.39/33.44/37.32 & 33.38& 0.9404/0.9770/0.9899 & 0.9691\\
    E-NeRV-Boost &0.75M/1.54M/3.07M & \textbf{32.61}/\textbf{37.19}/\textbf{40.07}& \textbf{36.62} &\textbf{0.9634}/\textbf{0.9857}/\textbf{0.9924} & \textbf{0.9805}\\
    \hline
    HNeRV \cite{chen2023hnerv}&0.75M/1.52M/3.03M & 30.55/35.13/38.15 & 34.61 & 0.9229/0.9731/0.9874 & 0.9611\\
    HNeRV-Boost & 0.76M/1.51M/3.08M & \textbf{35.09}/\textbf{38.52}/\textbf{41.09} & \textbf{38.23} & \textbf{0.9749}/\textbf{0.9882}/\textbf{0.9934} & \textbf{0.9855}\\
    % \hline
    % DivNeRV  &0.75M/1.52M/3.03M & 30.55/35.13/38.15 & 34.61 & 0.9229/0.9731/0.9874 & 0.9611\\
    % DivNeRV-Boost & 0.76M/1.51M/3.08M & \textbf{35.09}/\textbf{38.52}/\textbf{41.09} & \textbf{38.23} & \textbf{0.9749}/\textbf{0.9882}/\textbf{0.9934} & \textbf{0.9855}\\
    \hline
  \end{tabular}
\end{table*}

\begin{table*}
\setlength{\tabcolsep}{2pt}
\footnotesize
  \caption{Video regression results on the UVG dataset in PSNR and MS-SSIM.}
    \vspace{-0.25cm}
  \label{table:uvg_details_all}
  \centering
  \begin{tabular}{l|c|cccccccc|cccccccc}
    \hline
   \multirow{2}*{Model} &\multirow{2}*{Size} &  \multicolumn{8}{c|}{PSNR} &  \multicolumn{8}{c}{MS-SSIM} \\
    & & Beauty & Bosph. & Honey. & Jockey & Ready. & Shake. & Yacht. & Avg. & Beauty & Bosph. & Honey. & Jockey & Ready. & Shake. & Yacht. & Avg.\\
    \hline
    NeRV \cite{chen2021nerv}  &3.04M & 33.14 &	32.74 &	37.18 &	30.99 &	23.97 &	33.06 &	26.72 &	31.11 & 0.8918 & 0.9358 & 0.9806 & 0.8989 & 0.8426 & 0.9347 & 0.8712 & 0.9079 \\
    NeRV-Boost	  &3.06M & \textbf{33.55} & \textbf{34.51}&\textbf{39.04}&\textbf{32.82} &\textbf{26.08} & \textbf{34.54} & \textbf{28.76} & \textbf{32.76} & \textbf{0.8967} & \textbf{0.9480} & \textbf{0.9840} & \textbf{0.9174} & \textbf{0.8768} & \textbf{0.9458} & \textbf{0.8931} & \textbf{0.9231}\\
    E-NeRV \cite{li2022nerv}  &3.01M & 33.29 & 33.87 & 38.88 & 28.73 & 23.98 & 34.45 &  27.38 & 31.51 & 0.8933 & 0.9444 & 0.9843 & 0.8708 & 0.8449 & 0.9468 & 0.8842 & 0.9098 \\
    E-NeRV-Boost  &3.03M & \textbf{33.75} & \textbf{35.62} & \textbf{39.61} & \textbf{32.39} & \textbf{27.75} & \textbf{35.48} & \textbf{29.23} & \textbf{33.40} & \textbf{0.8987} & \textbf{0.9577} & \textbf{0.9854} & \textbf{0.9101} & \textbf{0.9057} & \textbf{0.9543} & \textbf{0.9015} & \textbf{0.9305}\\
    HNeRV \cite{chen2023hnerv} &3.05M &33.36 &	33.62 &	39.17 &	32.31 &	25.60 &	34.90 &	28.33 &	32.47 & 0.8907 & 0.9320 & 0.9843 & 0.8948 & 0.8490 & 0.9479 & 0.8642 & 0.9090 \\
    HNeRV-Boost	  &3.05M & \textbf{33.80} & \textbf{36.11} & \textbf{39.65} & \textbf{34.28} & \textbf{28.19} & \textbf{35.88} & \textbf{29.33} & \textbf{33.89} & \textbf{0.8996} & \textbf{0.9653} & \textbf{0.9854} & \textbf{0.9298} & \textbf{0.9139} & \textbf{0.958} & \textbf{0.9019} & \textbf{0.9363}\\
    \hline				
    NeRV \cite{chen2021nerv}  &5.07M & 33.62 & 34.32 & 38.32 & 32.86 & 25.67 & 34.24 & 28.06 & 32.44 & 0.8994 & 0.9528 & 0.9832 & 0.9230 & 0.8854 & 0.9488 & 0.9015 & 0.9277 \\
    NeRV-Boost	  &5.00M & \textbf{33.89} & \textbf{35.86} & \textbf{39.31} & \textbf{34.16} & \textbf{27.78} & \textbf{35.33} & \textbf{30.00} & \textbf{33.76} & \textbf{0.9020} & \textbf{0.9608} & \textbf{0.9846} & \textbf{0.9332} & \textbf{0.9072} & \textbf{0.9564} & \textbf{0.9160} & \textbf{0.9372}\\
    E-NeRV \cite{li2022nerv} &5.09M & 33.77 & 35.38 & 39.33 & 31.56 & 25.37 & 35.23 & 28.64 & 32.76& 0.9002 & 0.9596 & 0.9851 & 0.9050 & 0.8804 & 0.9561 & 0.9098 & 0.9280 \\
    E-NeRV-Boost  &5.01M & \textbf{34.02} & \textbf{36.79} & \textbf{39.71} & \textbf{33.90} & \textbf{29.29} & \textbf{36.20} & \textbf{30.24} & \textbf{34.31} & \textbf{0.9026} & \textbf{0.9669} & \textbf{0.9856} & \textbf{0.9287} & \textbf{0.9283} & \textbf{0.9626} & \textbf{0.9181} & \textbf{0.9418}\\
    HNeRV \cite{chen2023hnerv} &5.06M & 33.84 & 34.49 & 39.56 & 33.64 & 27.24 & 35.73 & 29.29 & 33.40& 0.8987 & 0.9430 & 0.9853 & 0.9114 & 0.8848 & 0.9588 & 0.8857 & 0.9240 \\
    HNeRV-Boost	  &5.01M & \textbf{34.14} & \textbf{37.87} & \textbf{39.74} & \textbf{35.84} & \textbf{30.36} & \textbf{36.71} & \textbf{30.77} & \textbf{35.06} & \textbf{0.9045} & \textbf{0.9764} & \textbf{0.9857} & \textbf{0.9467} & \textbf{0.9413} & \textbf{0.9675} & \textbf{0.9249} & \textbf{0.9496}\\
    \hline
    NeRV \cite{chen2021nerv}&10.10M & 34.10 & 36.52 & 39.35 & 35.37 & 28.10 & 35.82 & 30.11 & 34.20& \textbf{0.9088} & 0.9701 & 0.9852 & 0.9493 & 0.9302 & 0.9662 & 0.9354 & 0.9493 \\
    NeRV-Boost	  &10.08M & \textbf{34.17} & \textbf{37.77} & \textbf{39.65} & \textbf{36.23} & \textbf{30.25} & \textbf{36.81} & \textbf{32.06} & \textbf{35.28} & 0.9074 & \textbf{0.9749} & \textbf{0.9855} & \textbf{0.9525} & \textbf{0.9419} & \textbf{0.9703} & \textbf{0.9429} & \textbf{0.9536}\\
    E-NeRV \cite{li2022nerv} &10.16M & 34.18 & 37.31 & 39.70 & 34.62 & 28.27 & 36.50 & 30.36 & 34.42& 0.9065 & 0.9733 & 0.9858 & 0.9396 & 0.9297 & 0.9689 & 0.9361 & 0.9486 \\
    E-NeRV-Boost  &10.04M & \textbf{34.28} & \textbf{38.39} & \textbf{39.82} & \textbf{35.88} & \textbf{31.42} & \textbf{37.34} & \textbf{31.94} & \textbf{35.58} & \textbf{0.9065} & \textbf{0.9767} & \textbf{0.9859} & \textbf{0.9481} & \textbf{0.9515} & \textbf{0.9730} & \textbf{0.9389} & \textbf{0.9544}\\
    HNeRV \cite{chen2023hnerv} &10.07M & 34.22 & 37.27 & 39.73 & 34.59 & 29.59 & 36.82 & 30.70 & 34.70& 0.9053 & 0.9695 & 0.9857 & 0.9215 & 0.9255 & 0.9696 & 0.9134 & 0.9415 \\
    HNeRV-Boost	  &10.03M & \textbf{34.42} & \textbf{39.75} & \textbf{39.83} & \textbf{37.57} & \textbf{33.12} & \textbf{37.85} & \textbf{32.90} & \textbf{36.49} & \textbf{0.9096} & \textbf{0.984} & \textbf{0.9859} & \textbf{0.9617} & \textbf{0.9647} & \textbf{0.9768} & \textbf{0.9475} & \textbf{0.9615}\\
    \hline
    NeRV \cite{chen2021nerv}&15.09M & 34.36 & 37.66 & 39.59 & 36.55 & 29.81 & 36.86 & 31.43 & 35.18& \textbf{0.9170} & 0.9766 & \textbf{0.9857} & 0.9588 & \textbf{0.9509} & 0.9741 & 0.9508 & 0.9591 \\
    NeRV-Boost	  &15.04M & \textbf{34.47} & \textbf{38.87} & \textbf{39.67} & \textbf{37.35} & \textbf{30.87} & \textbf{37.37} & \textbf{33.00} & \textbf{35.94} & 0.9157 & \textbf{0.9803} & 0.9855 & \textbf{0.9622} & 0.9481 & \textbf{0.9742} & \textbf{0.9527} & \textbf{0.9598}\\
    E-NeRV \cite{li2022nerv} &15.02M & 34.34 & 38.41 & 39.78 &  35.98 & 29.90 & 37.32 & 31.52 & 35.32& \textbf{0.9111} & 0.9790 & 0.9860 & 0.9528 & 0.9492 & 0.9754 & 0.9491 & 0.9575 \\
    E-NeRV-Boost  &15.06M & \textbf{34.40} & \textbf{39.31} & \textbf{39.85} & \textbf{36.90} & \textbf{32.46} & \textbf{37.99} & \textbf{33.00} & \textbf{36.27} & 0.9089 & \textbf{0.9819} & \textbf{0.9860} & \textbf{0.9574} & \textbf{0.9598} & \textbf{0.9776} & \textbf{0.9496} & \textbf{0.9602}\\
    HNeRV \cite{chen2023hnerv} &15.02M & 34.37 & 38.40 & 39.81 & 35.76 & 31.02 & 37.00 & 31.82 & 35.45& 0.9079 & 0.9766 & 0.9859 & 0.9370 & 0.9435 & 0.9705 & 0.9295 & 0.9501 \\
    HNeRV-Boost	  &15.04M & \textbf{34.65} & \textbf{40.72} & \textbf{39.88} & \textbf{38.41} & \textbf{34.72} & \textbf{38.47} & \textbf{34.16} & \textbf{37.29} & \textbf{0.9176} & \textbf{0.9870} & \textbf{0.9861} & \textbf{0.9678} & \textbf{0.9739} & \textbf{0.9803} & \textbf{0.9578} & \textbf{0.9672}\\
    \hline
  \end{tabular}
\end{table*}

\begin{table}
\setlength{\tabcolsep}{2pt}
\scriptsize
  \caption{Video interpolation results on the UVG dataset in MS-SSIM.}
    \vspace{-0.25cm}
  \label{table:uvg_interpolation_msssim}
  \centering
  \begin{tabular}{l|cccccccc}
    \hline
    Model &  Beauty & Bosph. & Honey. & Jockey & Ready. & Shake. & Yacht. & Avg. \\
    \hline
    NeRV \cite{chen2021nerv} & \textbf{0.8696} & 0.9297 & 0.9797 & \textbf{0.7542} & \textbf{0.7070} & 0.9238 & 0.8578 & 0.8603 \\
    NeRV-Boost	&0.8673 & \textbf{0.9461} & \textbf{0.9835} & 0.7357 & 0.6970 & \textbf{0.9250} & \textbf{0.8708} & \textbf{0.8608} \\
    \hline
    E-NeRV \cite{li2022nerv} &0.8702 & 0.9383 & 0.9838 & \textbf{0.7536} & 0.7029 & \textbf{0.9288} & 0.8720 & 0.8642 \\
    E-NeRV-Boost&\textbf{0.8719} & \textbf{0.9525} & \textbf{0.9846} & 0.7476 & \textbf{0.7233} & 0.9272 & \textbf{0.8857} & \textbf{0.8704} \\
    \hline
    HNeRV \cite{chen2023hnerv} &0.8702 & 0.9379 & 0.9841 & 0.7677 & 0.7056 & 0.9263 & 0.8287 & 0.8601 \\
    HNeRV-Boost	&\textbf{0.8754} & \textbf{0.9664} & \textbf{0.9849} & \textbf{0.7836} & \textbf{0.7527} & \textbf{0.9284} & \textbf{0.8897} & \textbf{0.8830} \\
    \hline
  \end{tabular}
\end{table}

\begin{table}
\setlength{\tabcolsep}{2pt}
\footnotesize
  \caption{Encoding complexity comparison at resolution 1920$\times$1080 under video regression. The encoding time is evaluated by an NVIDIA V100 GPU.}
    \vspace{-0.25cm}
  \label{table:encoding_complexity}
  \centering
  \begin{tabular}{l|ccc}
   \hline
   Method   & Size  & Encoding time  & UVG PSNR\\
   \hline
   NeRV \cite{chen2021nerv} & 3.04M & 1h37m27s & 31.11\\
   NeRV-Boost  &3.06M & 3h54m15s &  32.76   \\
   \hline
   E-NeRV \cite{li2022nerv} & 3.01M & 2h47m42s & 31.51 \\
   E-NeRV-Boost  &3.03M & 5h14m23s  & 33.40 \\
   \hline
   HNeRV \cite{chen2023hnerv} & 3.05M & 7h3m34s &  32.47\\
   HNeRV-Boost & 3.06M & 13h51m0s   & 33.89 \\
   \hline
  \end{tabular}
\end{table}

\begin{table}
\setlength{\tabcolsep}{2pt}
\scriptsize
  \caption{Additional implicit models in video regression on the UVG dataset.}
    \vspace{-0.25cm}
  \label{table:baselines}
  \centering
  \begin{tabular}{c|cccc}
    \hline
    Model &  Size & Resolution & PSNR Avg. & MS-SSIM Avg. \\
    \hline
    DNeRV \cite{he2023towards}     & 8.02M &	1024$\times$1920 &	34.73 & 0.9498  \\
    DNeRV-Boost	& 8.08M  &	1024$\times$1920 & \textbf{35.50} & \textbf{0.9548} \\
    \hline
    DivNeRV \cite{zhao2023dnerv} & 10.08M &	1080$\times$1920 &	33.93 & 0.9316 \\
    DivNeRV-Boost & 10.02M &	1080$\times$1920 &	\textbf{35.35} & \textbf{0.9521} \\
    \hline
  \end{tabular}
  \vspace{-0.45cm}
\end{table}

\begin{table}
\footnotesize
  \caption{Ablation study of different boosting components on the Buuny video with 3M model size and 300 training epochs.}
  \vspace{0.05cm}
  \label{table:ablation_components}
  \centering
  \begin{tabular}{c|ccc|c}
    \hline
    Model & TAT & SNeRV & Improved loss & PSNR \\
    \hline
    \multirow{4}*{NeRV \cite{chen2021nerv}} & &  &  & 31.84  \\
    & \checkmark &  &  & 33.50  \\
     & \checkmark & \checkmark & & 36.28  \\
    & \checkmark & \checkmark & \checkmark & \textbf{37.25}  \\
    \hline
    \multirow{4}*{E-NeRV \cite{li2022nerv}} & &  &  & 37.32  \\
    & \checkmark &  &  & 38.01 \\
     & \checkmark & \checkmark & &  39.34 \\
    & \checkmark & \checkmark & \checkmark & \textbf{40.07}  \\
    \hline
    \multirow{4}*{HNeRV \cite{chen2023hnerv}} & &  &  & 38.15  \\
    & \checkmark &  &  & 39.90  \\
     & \checkmark & \checkmark & & 40.19 \\
    & \checkmark & \checkmark & \checkmark & \textbf{41.09}  \\
    % \hline
    % \multirow{4}*{DivNeRV \cite{zhao2023dnerv}} & &  &  & 34.29  \\
    % & \checkmark &  &  & xx  \\
    %  & \checkmark & \checkmark & & 38.78 \\
    % & \checkmark & \checkmark & \checkmark & \textbf{39.69}  \\
    \hline
  \end{tabular}
\end{table}

\begin{table*}[t]
\setlength{\tabcolsep}{2pt}
\scriptsize
  \caption{Video inpainting results on the DAVIS validation dataset in MS-SSIM.}
    \vspace{-0.25cm}
  \label{table:davis_details_msssim}
  \centering
  \begin{tabular}{l|cccccc|cccccc}
    \hline
    \multirow{2}*{Video} &  \multicolumn{6}{c|}{Mask-S} & \multicolumn{6}{c}{Mask-C}  \\
    &  NeRV & NeRV-Boost  & E-NeRV & E-NeRV-Boost & HNeRV & HNeRV-Boost  & NeRV & NeRV-Boost  & E-NeRV & E-NeRV-Boost & HNeRV & HNeRV-Boost  \\
    \hline
Blackswan	         &0.8687 &\textbf{0.9234} &	0.9216 &\textbf{0.9362} &	0.9002 &\textbf{0.9626} &	0.8314 &\textbf{0.8869} &	0.8857 &\textbf{0.9021} &	0.8736 &\textbf{0.9292}\\
Bmx-trees	         &0.8421 &\textbf{0.9124} &	0.8905 &\textbf{0.9239} &	0.8676 &\textbf{0.9519} &	0.8091 &\textbf{0.8780} &	0.8537 &\textbf{0.8992} &	0.8308 &\textbf{0.9062}\\
Breakdance	         &0.9347 &\textbf{0.9497} &	0.9571 &\textbf{0.9665} &	0.9184 &\textbf{0.9791} &	0.8857 &\textbf{0.9016} &	0.9083 &\textbf{0.9152} &	0.9172 &\textbf{0.9257}\\
Camel	             &0.8178 &\textbf{0.8699} &	0.8790 &\textbf{0.9011} &	0.8577 &\textbf{0.9480} &	0.7784 &\textbf{0.8319} &	0.8397 &\textbf{0.8689} &	0.8186 &\textbf{0.9009}\\
Car-roundabout	     &0.8741 &\textbf{0.9301} &	0.9151 &\textbf{0.9407} &	0.9194 &\textbf{0.9596} &	0.8323 &\textbf{0.8868} &	0.8752 &\textbf{0.9047} &	0.8798 &\textbf{0.9149}\\
Car-shadow	         &0.9042 &\textbf{0.9555} &	0.9515 &\textbf{0.9558} &	0.9367 &\textbf{0.9685} &	0.8673 &\textbf{0.9132} &	0.9059 &\textbf{0.9158} &	0.7635 &\textbf{0.9256}\\
Cows	             &0.7408 &\textbf{0.8445} &	0.8346 &\textbf{0.9026} &	0.7983 &\textbf{0.9408} &	0.6957 &\textbf{0.8041} &	0.7917 &\textbf{0.8564} &	0.7627 &\textbf{0.9022}\\
Dance-twirl	         &0.8296 &\textbf{0.8866} &	0.8682 &\textbf{0.8959} &	0.8544 &\textbf{0.9223} &	0.7891 &\textbf{0.8457} &	0.8265 &\textbf{0.8569} &	0.8074 &\textbf{0.8842}\\
Dog	                 &0.8770 &\textbf{0.9321} &	0.9382 &\textbf{0.9455} &	0.8297 &\textbf{0.9636} &	0.8383 &\textbf{0.8921} &	0.8995 &\textbf{0.9107} &	0.8655 &\textbf{0.9212}\\
Drift-chicane	     &0.9747 &\textbf{0.9870} &	0.9870 &\textbf{0.9899} &	0.9806 &\textbf{0.9924} &	0.9360 &\textbf{0.9435} &	0.9539 &\textbf{0.9583} &	0.9360 &\textbf{0.9484}\\
Drift-straight	     &0.9134 &\textbf{0.9612} &	0.9499 &\textbf{0.9670} &	0.9261 &\textbf{0.9820} &	0.8662 &\textbf{0.9160} &	0.9040 &\textbf{0.9260} &	0.8452 &\textbf{0.9324}\\
Goat	             &0.8116 &\textbf{0.8661} &	0.8585 &\textbf{0.8885} &	0.8699 &\textbf{0.9471} &	0.7709 &\textbf{0.8239} &	0.8187 &\textbf{0.8462} &	0.8270 &\textbf{0.8972}\\
Horsejump-high	     &0.8861 &\textbf{0.9306} &	0.9238 &\textbf{0.9316} &	0.9079 &\textbf{0.9416} &	0.8462 &\textbf{0.8905} &	0.8847 &\textbf{0.8937} &	0.8641 &\textbf{0.8997}\\
Kite-surf	         &0.9048 &\textbf{0.9511} &	0.9504 &\textbf{0.9592} &	0.9226 &\textbf{0.9725} &	0.8676 &\textbf{0.9136} &	0.9146 &\textbf{0.9258} &	0.8781 &\textbf{0.9302}\\
Libby	             &0.8895 &\textbf{0.9504} &	0.9355 &\textbf{0.9576} &	0.7921 &\textbf{0.9756} &	0.8539 &\textbf{0.9123} &	0.8956 &\textbf{0.9232} &	0.7373 &\textbf{0.9372}\\
Motocross-jump	     &0.9251 &\textbf{0.9702} &	0.9601 &\textbf{0.9683} &	0.8524 &\textbf{0.9747} &	0.8930 &\textbf{0.9360} &	0.9271 &\textbf{0.9314} &	0.8319 &\textbf{0.9403}\\
Paragliding-launch 	 &0.8615 &\textbf{0.9063} &	0.8999 &\textbf{0.9214} &	0.8920 &\textbf{0.9416} &	0.8366 &\textbf{0.8800} &	0.8769 &\textbf{0.8970} &	0.8626 &\textbf{0.9158}\\
Parkour	             &0.8152 &\textbf{0.8840} &	0.8565 &\textbf{0.8888} &	0.8399 &\textbf{0.9142} &	0.7780 &\textbf{0.8493} &	0.8203 &\textbf{0.8554} &	0.7998 &\textbf{0.8706}\\
Scooter-black	     &0.8894 &\textbf{0.9422} &	0.9406 &\textbf{0.9521} &	0.9349 &\textbf{0.9647} &	0.8480 &\textbf{0.8960} &	0.8925 &\textbf{0.9058} &	0.8849 &\textbf{0.9145}\\
Soapbox	             &0.8689 &\textbf{0.9159} &	0.8960 &\textbf{0.9231} &	0.8907 &\textbf{0.9392} &	0.8334 &\textbf{0.8801} &	0.8570 &\textbf{0.8877} &	0.8483 &\textbf{0.8917}\\
\hline
Average	             &0.8715 &\textbf{0.9235} &	0.9157 &\textbf{0.9358} &	0.8846 &\textbf{0.9571} &	0.8329 &\textbf{0.8841} &	0.8766 &\textbf{0.8990} &	0.8417 &\textbf{0.9144}\\
    \hline
  \end{tabular}
\end{table*}

\begin{table*}
\footnotesize
  \caption{Details of each module. Conv(input channels, output channels, kernel size, stride). PixelShuffle(upscale factor). $r$ denotes the reduction factor of the channel width.}
  \label{table:sub_modules}
  \vspace{-0.3cm}
  \centering
  \begin{tabular}{c|c|c|c|c}
    \hline
  \multicolumn{2}{c|}{TAT($C$)} & TAT Resblock($C$) & SNeRV($C$, $r$, $k$, $s$) & Sinusoidal E-NeRV($C$, $r$, $k$, $s$) \\
    \hline
   Conv(32, 32, 1, 1) & Conv(32, 32, 1, 1) & TAT($C$) & Conv($C$, $C*s*s/r$, $k$, 1) & Conv($C$, $C*s*s/4$, $k$, 1) \\
   ReLU() & ReLU() & Conv($C$, $C$, 3, 1) & \multirow{2}*{$\left\{\begin{aligned} {\rm \text{Identity}}(), s=1 \\
   {\rm \text{PixelShuffle}}(s), s>1 \end{aligned}\right.$}  & PixelShuffle$(s)$\\
  Conv(32, $C$, 1, 1)&  Conv(32, $C$, 1, 1) &  GELU() & & Conv($C/4$, $C/r$, 3, 1)\\
   &   & TAT($C$) &  & SINE() \\
   &  & Conv($C$, $C$, 3, 1) & SINE()  & \\
  &  &  Skip connection &  & \\
    \hline
  \end{tabular}
\end{table*}

\begin{table*}
\footnotesize
  \caption{Details of conditional decoders in different implicit models. $C_{i+1}=\lfloor C_i/r, C_{min} \rfloor$, where the minimum channel width $C_{min}$ is set as 12.}
  \label{table:conditional_decoder}
    \vspace{-0.3cm}
  \centering
  \begin{tabular}{c|c|c|c}
    \hline
   $z_t$ generation network & NeRV-Boost & E-NeRV-Boost & HNeRV-Boost \\
    \hline
    Reshape            & SNeRV($C_1$, 1, 3, $s_1$)  & Sinusoidal E-NeRV($C_1$, 1/3, 3, $s_1$) & SNeRV(16, 16/$C_1$, 1, 1) \\
    Conv(160, 64, 1, 1)& TAT($C_1$)                 & TAT($C_1*3$)                        & TAT($C_1$) \\
    SINE()             & SNeRV($C_1$, 2, 3, $s_2$)  & SNeRV($C_1*3$, 2, 3, $s_2$)         & SNeRV($C_1$, 1.2, 3, $s_1$) \\
    Conv(64, 32, 1, 1) & TAT($C_2$)                 & TAT($C_2$)                          & TAT($C_2$) \\
    SINE()             & SNeRV($C_2$, 2, 3, $s_3$)  & SNeRV($C_2$, 2, 3, $s_3$)           & SNeRV($C_2$, 1.2, 3, $s_2$) \\
                       & TAT($C_3$)                 & TAT($C_3$)                          & TAT($C_3$)  \\
                       & SNeRV($C_3$, 1, 3, 1)      & SNeRV($C_3$, 1, 3, 1)               & SNeRV($C_3$, 1.2, 3, $s_3$) \\
                       & TAT($C_3$)                 & TAT($C_3$)                          & TAT($C_4$) \\
                       & SNeRV($C_3$, 2, 3, $s_4$)  & SNeRV($C_3$, 2, 3, $s_4$)           & SNeRV($C_4$, 1, 3, 1) \\
                       & TAT($C_4$)                 & TAT($C_4$)                          & TAT($C_4$) \\
                       & SNeRV($C_4$, 1, 3, 1)      & SNeRV($C_4$, 1, 3, 1)               & SNeRV($C_4$, 1.2, 3, $s_4$) \\
                       & TAT($C_4$)                 & TAT($C_4$)                          & TAT($C_5$) \\
                       & SNeRV($C_4$, 2, 3, $s_5$)  & SNeRV($C_4$, 2, 3, $s_5$)           & SNeRV($C_5$, 1, 3, 1) \\
                       & TAT($C_5$)                 & TAT($C_5$)                          & TAT($C_5$) \\
                       & SNeRV($C_5$, 1, 3, 1)      & SNeRV($C_5$, 1, 3, 1)               & SNeRV($C_5$, 1.2, 3, $s_5$)  \\
                       & TAT($C_5$)                 & TAT($C_5$)                          & TAT($C_6$) \\
                       & Conv($C_5$, 3, 1, 1)       & Conv($C_5$, 3, 1, 1)                & SNeRV($C_6$, 1, 3, 1)  \\
                       &                            &                                     & TAT($C_6$)  \\
                       &                            &                                     & Conv($C_6$, 3, 1, 1) \\
    \hline
  \end{tabular}
\end{table*}

% \section{Limitations}

\end{document}